\documentclass[pdflatex,sn-mathphys-num]{sn-jnl}

\usepackage{graphicx}%
\usepackage{multirow}%
\usepackage{amsmath,amssymb,amsfonts}%
\usepackage{amsthm}%
\usepackage{mathrsfs}%
\usepackage[title]{appendix}%
\usepackage{xcolor}%
\usepackage{textcomp}%
\usepackage{manyfoot}%
\usepackage{booktabs}%
\usepackage{algorithm}%
\usepackage{algorithmicx}%
\usepackage{algpseudocode}%
\usepackage{listings}%

\theoremstyle{thmstyleone}%
\theoremstyle{thmstyletwo}%

\theoremstyle{thmstylethree}%

\begin{document}

\title[Ion acceleration from micrometric targets immersed in an intense laser field]{Ion acceleration from micrometric targets immersed in an intense laser field}


\author[1,2]{\fnm{Michal} \sur{Elkind}}
\author[1,2]{\fnm{Noam} \sur{Popper}}
\author[1,2]{\fnm{Itamar} \sur{Cohen}}
\author[1,2]{\fnm{Aviv} \sur{Levinson}}
\author[1,2]{\fnm{Nitzan} \sur{Alaluf}}
\author[1,2]{\fnm{Assaf} \sur{Levanon}}
\author*[1,2]{\fnm{Ishay} \sur{Pomerantz}}

\affil[1]{\orgdiv{The School of Physics and Astronomy}, \orgname{Tel Aviv University}, \orgaddress{\city{Tel Aviv}, \postcode{69978}, \country{Israel}}}

\affil[2]{\orgdiv{Center for Light-Matter Interaction}, \orgname{Tel Aviv University}, \orgaddress{\city{Tel Aviv}, \postcode{69978}, \country{Israel}}}


\abstract{We report on an experimental study of proton acceleration by intense laser irradiation of micrometric bar targets,
whose dimensions are transversely immersed in the laser focal volume and are longitudinally smaller than half its wavelength.
With only 120 mJ of laser energy, we recorded proton energies in excess of 6~MeV,  
 three times higher than those achieved with flat-foil irradiation using similar pulse energies.
3D particle-in-cell simulations revealed that the efficient energy transfer from the diffracted laser fields to electrons on both sides of the target, combined with its reduced surface area, results in a thicker electron sheath and higher acceleration gradients.
We demonstrated numerically how this technique opens up the possibility of laser-ion acceleration in a cascaded manner, allowing manipulation of the ion spectrum by optical means. 
}




\maketitle

The interaction of an intense laser pulse with matter results in the emission of multiple forms of radiation,
including x-rays, electrons, ions, and positrons. 
This general observation has motivated three decades of research on laser-based particle acceleration. 
The prospect of accelerating ions on a compact scale to MeV energies and beyond has potential for many applications, including radiography of transient phenomena and strong electromagnetic fields \cite{RevModPhys.95.045007}, for the ion fast-ignition approach to fusion energy \cite{fernandez2014fast}, and in generating neutron beams \cite{roth2013bright} for non-destructive testing \cite{kishon2019laser}. It has particularly promising benefits in radiation therapy \cite{ledingham2014towards}
because of the enormous cost of ion radiation therapy based on current technology, which limits the use of this treatment. 
A fundamental requirement for this application is the acceleration of ions to energies sufficiently high to penetrate human tissue and reach any tumor, which in the case of protons is about 250 MeV \cite{kumada2020accelerator}. 

The target normal sheath acceleration (TNSA) mechanism of ions \cite{passoni2010target} 
has been extensively studied in dozens of different experimental scenarios.
TNSA relies on high-magnitude electric fields that form between an irradiated target
and the electron sheath that develops around it to accelerate ions from surface contaminants.
The general phenomenology is that laser systems that deliver higher pulse energies and shorter pulse durations are able to accelerate ions from flat-foil targets to higher energies \cite{zimmer2021analysis}. 
For example, the current record of 150 MeV \cite{ziegler2024laser} has been achieved using ultrashort laser pulses with 22 J of energy,
whereas 0.12 J laser pulses, like the one used for the results presented here, typically reach maximum proton energies of 2 MeV \cite{zimmer2021analysis,gershuni2019gatling}.
Prospective applications suitable for such lower energies include
radiopharmaceuticals generation either directly \cite{maffini2023laser}
or through the products of secondary nuclear reactions  \cite{PhysRevResearch.7.013230,batani2025generation}, and materials analysis techniques such as fast neutron activation \cite{PhysRevApplied.19.044020} and laser-driven particle-induced x-ray emission \cite{mirani2021integrated}.

Advanced target designs aimed to enhance TNSA
via controlled pre-expansion, localized field enhancement, or increased heating of target electrons include foam layers  \cite{Sgattoni2012,prencipe2021efficient,Jiang2018}, sub-micron gratings \cite{ceccotti2013evidence,sgattoni2015laser,blanco2017table}, micron-scale channels \cite{Lezhnin2022,Zou2019,Yang2018}, and micro- or nano-structured wire arrays \cite{Khaghani2017,Frontiers2023}.

In TNSA, both the lateral and longitudinal (thickness) dimensions of an irradiated target affect the characteristics of the emitted ions.
It is generally established that ion energies rise when the target thickness is of the order of the laser wavelength or smaller
\cite{zimmer2021analysis}. The caveat is that very thin targets would remain intact throughout the interaction to emit ions only when the laser pulse contrast is sufficiently high \cite{keppler2022intensity}.
Limiting the transverse size of thick target foils down to 10s of micrometers was observed to enhance
the accelerating gradient of the electron sheath because of electrons refluxing from the edges of the target 
\cite{buffechoux2010hot,kraft2018first,zeil2014robust,toncian2011optimal,tresca2011controlling,fang2016different}. 
But in this case, the resulting ion energies still fall short of those produced using sub-wavelength thin foils irradiated with the same laser pulse parameters \cite{green2014high,hornung2020enhancement,busold2014construction}.

Here we report on irradiation experiments of single formations immersed in the focal volume of an intense laser pulse and thinner than half its wavelength.
These experiments resulted in proton energies higher than those achieved with any other laser-based method with a similar pulse energy. 
We show that this effect is not a mere combination of the two aforementioned observations about the dimensions of the target but rather a manifestation of a different and more efficient ability of the laser fields to transfer energy to the electrons that form the sheath. 
Hints of this dynamics could be gleaned from several experimental results showing enhanced proton emission from irradiated surfaces covered with nanometric or micrometric structures \cite{Zigler2013,Floquet2013,margarone2012laser,curtis2018micro}.

Irradiation of truly isolated targets fully contained within the laser focal volume is mechanically challenging
and was demonstrated only by using a Paul trap to levitate single micrometer-scale plastic spheres \cite{Ostermayr2016spheres,hilz2018isolated}. These experiments did not result in higher proton energies than those obtained with solid foils but the emitted protons featured a reduced energy bandwidth. 

We studied the emission of protons from irradiated micrometric bar targets (µ-bars) made of gold.
In a previous work we have shown  that the interaction of an intense laser pulse with such µ-bars results in the emission of two beams of MeV-level electrons with a narrow opening angle \cite{elkind2023intense}.
Numerical simulations revealed that these beams consist of trains of attosecond-duration electron bunches,
emitted because of the diffraction of the fields around the target.
These bunches interact with the laser field over large distances in vacuum, resulting in non-trivial dependence on the focusing geometry \cite{de2024unforeseen}.

\section*{Results}
\begin{figure}[!htb]%
\centering
\includegraphics[width=0.75\textwidth]{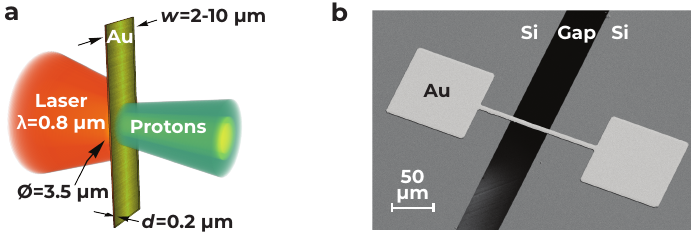}
\caption{The irradiation setup. (a) Illustration of the irradiation geometry.
(b) SEM image of a $w$~=~5~µm wide µ-bar suspended over a gap in a Si substrate.}
\label{fig:exp-setup}
\end{figure}

The irradiation geometry (see Methods) is illustrated in Fig.~\ref{fig:exp-setup}a. 
We irradiated $d$ = 0.2 µm thick, $w$ = 2--10 µm wide µ-bars using
27-fs long p-polarized laser pulses with 120 mJ of energy. These pulses were focused to a 3.5 µm diameter spot, corresponding to a normalized laser amplitude  of $a_0$~=~4.6.
A scanning electron microscope image of one such target is shown in Fig.~\ref{fig:exp-setup}b. See the Methods section for details about how the µ-bars were fabricated.

\begin{figure}[!htb]%
\centering
\includegraphics[width=0.68\textwidth]{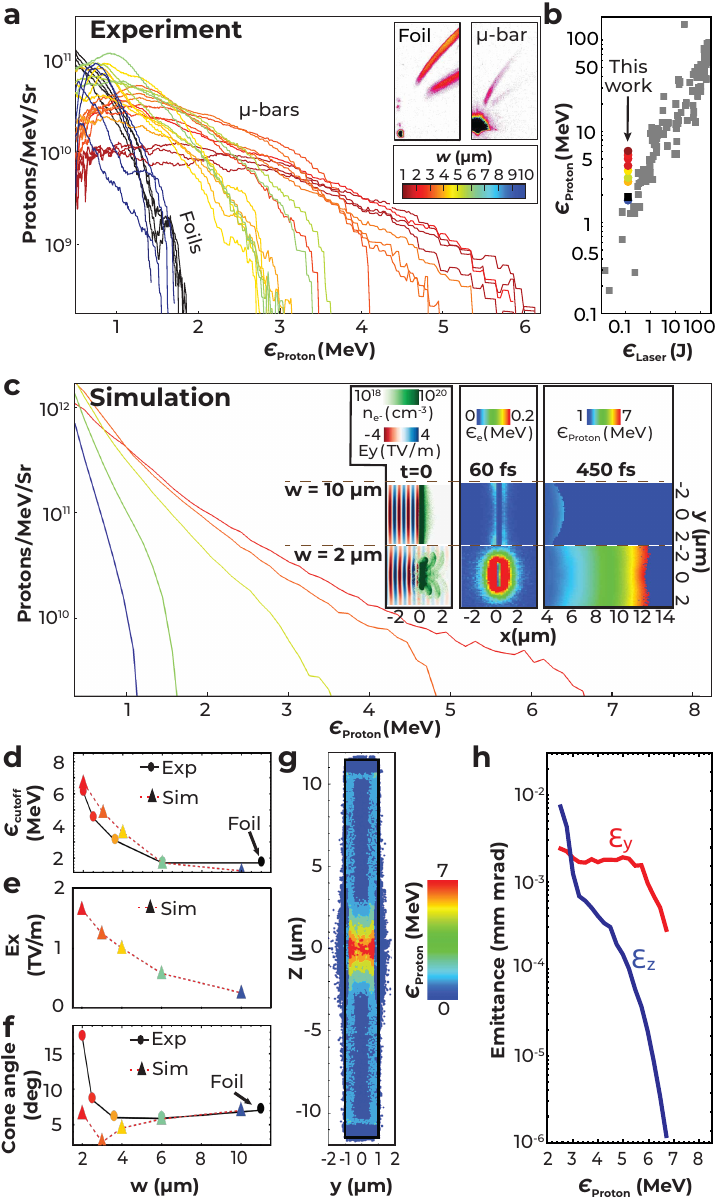}
\caption{
The effect of the µ-bar width.
(a) Experimentally measured differential proton energy spectra emitted from $d$~=~0.2~µm thick Au foils (black) and µ-bar targets (color scale). Example raw TPIS traces resulting from irradiation of a foil target and a $w$~=~1.8~µm wide µ-bar are shown as insets. 
(b) Compilation of the proton cutoff energies from experiments of various laser and target parameters, as a function of the laser pulse energy (gray) adapted from Ref.~\cite{zimmer2021analysis}.  The results of this work are shown with the µ-bar width indicated in color.
 (c) Simulated differential proton energy spectra for the same experimental  parameters as in (a), overlaid with  snapshots of the transverse electric field (blue-to-red), electron density (green) and space averaged electron and proton energy (color scale).
 (d-f) Measured and simulated proton energy cutoffs, peak-values of the electric field  in the sheath, and the cone angles of protons emitted with $E > 0.5$~MeV plotted vs. $w$.
(g) Energy resolved virtual source distribution of the protons and (h) the RMS of the emittance for a $d$~=~0.2~µm, $w$~=~2~µm µ-bar.}
\label{fig:proton spectra}
\end{figure}

The energy spectra of protons emitted at the laser propagation direction were measured using a Thomson parabola type ion spectrometer (TPIS) \cite{morrison2011design} described in detail in the Methods section. The resulting differential proton spectra for flat foils (black) and µ-bars ($w$ in a color scale) are shown in Fig.~\ref{fig:proton spectra}a. Each curve represents the result of a single irradiation experiment.
Two raw spectrograms are shown in insets for an irradiated reference 0.2 µm thick Au foil and a $d$ = 0.2 µm thick, $w$ = 1.8 µm wide µ-bar.
The increased signal around the zero-point results
from the electron jets that we studied in Ref. \cite{elkind2023intense}, impinging on the spectrometer wall. 
The irradiation of narrower µ-bars feature 
higher proton cutoff energies, reaching beyond 6 MeV for $w$~=~2~µm.

Fig.~\ref{fig:proton spectra}b presents a compilation of the proton cutoff energies in laser acceleration experiments as a function of the laser pulse energy,
adapted from Ref. \cite{zimmer2021analysis} and references therein. 
Also shown are the results of this study for the irradiation of µ-bar targets, where $w$ is indicated with the same color scale as in Fig.~\ref{fig:proton spectra}a. 
The figure shows that in a common flat-foil irradiation scenario (gray squares), a laser pulse energy higher than 1 J is required to accelerate protons to the same cutoff energies as we achieved using 120~mJ only.

The underlying dynamics were revealed by 3D particle-in-cell (PIC)  simulations using the EPOCH \cite{arber2015contemporary}  code (see the Methods section for details). 
In these simulations, $d$ = 0.2 µm thick µ-bars of various widths were irradiated with p-polarized 800-nm wavelength laser pulses having a 30 fs (FWHM) wide Gaussian temporal profile and 120 mJ of energy.
The laser pulses were focused to a spot size of 3.5 µm (FWHM), yielding a normalized laser amplitude of $a_0$~=~4.6. The simulation results are presented in Fig.~\ref{fig:proton spectra}c.  
The differential proton energy spectra for irradiated µ-bar targets in the same parameter range as in Fig.~\ref{fig:proton spectra}a are shown with the same color scale.
Overlaid are snapshots taken at $t$~=~0,~60,~and~450~fs, for the cases of
$w$~=~10~µm and 2~µm wide µ-bars.
t = 0 represents the instant in which the peak of the laser field impinges on the µ-bar.
The transverse component of the laser field (E$_y$) is shown in a red-to-blue color scale, with the electron density superimposed in a green color scale.
The $w$~=~2~µm µ-bar is narrower than the laser focus and therefore is transversely immersed in the focal volume. 

The experimentally observed increase in the proton cutoff energy for narrower µ-bars is captured by the simulation (Fig.~\ref{fig:proton spectra}d).
Fig.~\ref{fig:proton spectra}e shows how the increased proton cutoff energies for narrower µ-bars are correlated with
the sheath field amplitude (snapshots taken at $t$ = 60 fs). 
The cone angles of $E >$ 0.5 MeV proton beams emitted from $d$~=~0.2~µm µ-bar targets are plotted in Fig.~\ref{fig:proton spectra}f as a function of $w$. 
Both the experiment and the simulation feature an increased divergence of the proton beam for the narrowest ($w$~=~2~µm) targets. 
This geometric effect occurs when both $d$ and $w$ are smaller than the
sheath scale length, which is on the order of microns \cite{jackel2010all}, 
so the sheath no longer maintains the target's aspect ratio.

For narrow µ-bars, the measured proton energy spectra feature flattening in the low-energy region that is not captured by simulations. Simulating the low-energy tail of laser-accelerated ions remains a challenge \cite{macchi2013advanced}, as it often requires extremely high numbers of computational particles 
to accurately represent a realistic plasma scale-length, when the spatial resolution is limited by computational constraints \cite{lecz2013target}.
This spectral flattening may stem from several numerically unresolved physical mechanisms, including fast electrons refluxing through the target that induces oscillating charge separation at the rear surface \cite{dubois2017origins}; the emergence of  ion acceleration mechanisms other than TNSA \cite{fang2016combined}; and high sensitivity to the detailed density profiles of surface contaminants \cite{robinson2006effect}.

Strong sheath fields close to the target may modify the proton trajectories following their emission. This affects the ability of a distant observer to determine the spread of their point of origin (their "source size"). 
To account for this effect, we obtained
energy resolved “virtual” source distributions of the proton beam \cite{borghesi2004multi}  by projecting the proton angle at the end of the acceleration phase back to the target plane. 
These are shown  in Fig.~\ref{fig:proton spectra}g for the case of a $d$~=~0.2~µm, $w$~=~2~µm µ-bar, 
color-coded by the final proton energy, and features a source size smaller than 4 µm$^2$ for $E >$ 4 MeV protons.
The normalized r.m.s. values of the two transverse components of the proton beam emittance are shown in Fig.~\ref{fig:proton spectra}h. 
These were evaluated as
$\mathcal{E}_{y/z}=(p/mc)\sigma_{y/z}\sigma_{{y'/z'}}$ 
where $\sigma_{y/z}, \sigma_{y'/z'}$ are the r.m.s. values of the source beam width and divergence angle in the $y$ and $z$ directions \cite{roth2017ion}.

\begin{figure}[!htb]
\centering
\includegraphics[width=0.65\textwidth]{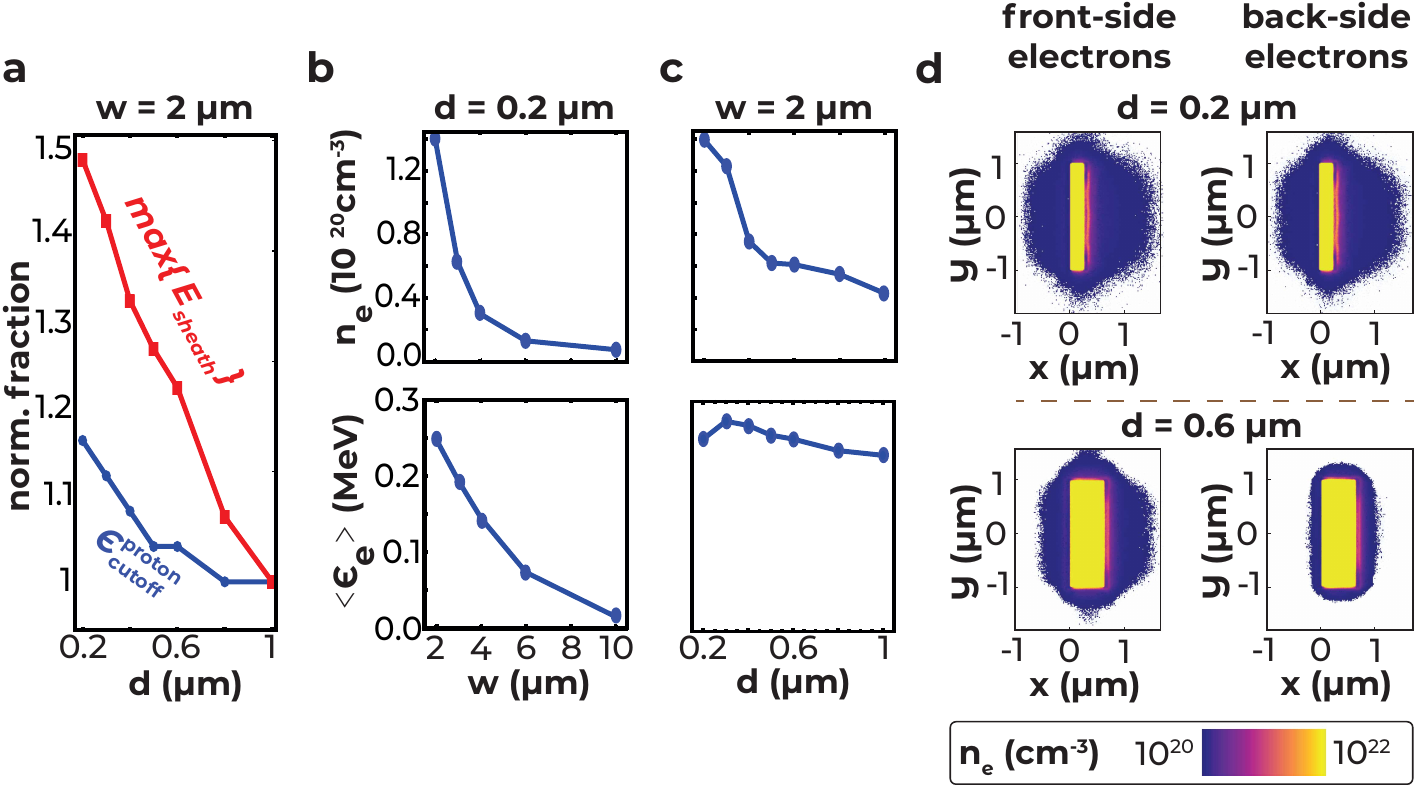}
\caption{The effect of the µ-bar thickness.
(a) Proton cutoff energies (blue) and the maximal amplitude of the sheath field (sampled at $t$~=~60~fs, red).
(b) The density and average energy of electrons in the sheath sampled at $t$~=~60~fs for $d$~=~0.2~µm and $w$~=~2--10~µm µ-bars.
(c) Same as (b) for $w$~=~2~µm and $d$~=~0.2--1.0~µm µ-bars.
(d) Density distributions of the electron sheath forming at t~=~60~fs around $w$~=~2~µm, $d$~=~0.2~µm (top) and 0.6~µm (bottom) µ-bars.
Electrons originating from the front side or the back side of the target are shown on the left and right respectively.}
\label{fig:pic}
\end{figure}

So far we discussed how the lateral dimension $w$ of the target plays a role in increasing the emitted proton energies. 
The effect of the target thickness $d$ on the proton maximal energy is
demonstrated in Fig.~\ref{fig:pic}a; a sharp increase of up to 15\% in the proton cutoff energy (blue curve) emerge for $d<\lambda/2$ targets.
This increase in energy is correlated with an increase in the peak value of the sheath electric field (red curve).
Some properties of TNSA may be obtained using a simple self-similar isothermal fluid model \cite{mora2003plasma,fuchs2006laser} in which the proton cutoff energy is given by
$\epsilon_{\rm{cutoff}} = 2T_e ln[(\tau+\sqrt{\tau^2+1})^2]$
with $\tau = \omega_{pi} t_{acc}/\sqrt{2e}$.
Here $T_e$ is the temperature of the hot electron population, 
$t_{acc}$ is an effective acceleration time, and
$\omega_{pi}\sim \sqrt{n_e}$ is the ion plasma frequency.
The strong dependence of $\epsilon_{\rm{cutoff}}$ on $T_e\sim\langle\epsilon_e\rangle$ is observed in Fig.~\ref{fig:pic}b where the sheath density and average electron energy were sampled at $t$~=~60~fs, 1-µm behind the rear side of the target.
For $d$~=~0.2~µm µ-bars, smaller values of $w$ result in higher electron energies that can account for the higher energy cutoff.
However, when reducing $d$ for fixed $w$~=~2~µm µ-bars (Fig.~\ref{fig:pic}c),
the electron energies do not increase
and the source of the increased proton cutoff energies is found to be the rising sheath density.
To identify the origin of this thicker sheath, we separated the electron population according to the surface from which they originated. 
Fig.~\ref{fig:pic}d presents sheath density distributions forming around $w$~=~2~µm µ-bars at t~=~60~fs.
Electrons that initially covered the plasma gradient at the front ($x < 0$) of the target are shown on the left, and those that originated from the back side ($x-d > 0$) are shown on the right.
For a target thinner than half the laser wavelength (top, $d$ = 0.2 µm), 
the laser field is sufficient to pull electrons from the rear of the target and 
the sheath is observed to be a mixture of front and back electrons.
However, when the µ-bar is thicker than half the laser wavelength (bottom, $d$ = 0.6 µm), the sheath is observed to consist of front-side electrons only.

The inset in Fig. \ref{fig:proton spectra}(c) features Coulomb expansion of the electron sheath for a $w$~=~2~µm, $d$~=~0.2~µm µ-bar. The drop in proton energies for thicker targets in Fig. \ref{fig:pic}(a) indicates that the higher proton energies are indeed due to enhanced TNSA rather than to Coulomb explosion (i.e., not from the target ions accelerating themselves by mutual repulsion), since in the latter case proton energies would rise due to the increased stored charge.

\section*{Discussion}
\subsubsection*{Source properties}
One of the strong suites of using µ-bar targets for TNSA we identified is the small virtual source size of the protons.  
This property sets a limit on the spatial resolution when performing proton radiography,
a method used for a range of basic research \cite{RevModPhys.95.045007} and medical applications \cite{johnson2017review}.
In TNSA off planar foils the virtual source size is 
of the order of 10 µm \cite{borghesi2004multi,wang2015large,li2021influence},
while for µ-bars (Fig.~\ref{fig:proton spectra}g)
it was found to be smaller than 1~µm for the high energy part of the proton beam.
This advantage is further highlighted by the low transverse $z$ component of the proton beam
emittance, plotted in Fig.~\ref{fig:proton spectra}h, 
which drops well below the values typical to planar foil targets ($\sim$10$^{-3}$~mm~mrad \cite{roth2017ion}).

Compared to the common irradiation scenario of flat foils, 
the widest µ-bars we irradiated ($w$~=~6~µm), 
produce the same absolute spectrum of protons above $E > 1.5$~MeV (Fig.~\ref{fig:proton spectra}a).
Key properties of the resulting proton beams for the cases of foil targets and $w$~=~2~µm  µ-bars (the narrowest we irradiated) are given in Table~1.
For µ-bars, the 32\% lower proton flux at the laser direction is nearly balanced by the increased emission cone-angle  (Fig.~\ref{fig:proton spectra}f), 
yielding similar total proton numbers in both cases. However, the higher proton energies in the µ-bar case corresponds to a $\sim$4 times higher coupling efficiency of laser energy to the proton beam, compared to that of conventional foil targets.

\begin{table}
    \centering
    \begin{tabular}{|l|l|l|l|} 
    \hline
        & no. protons/sr  & no. protons & conversion eff.  \\
    \hline&&&\\[-1em]
         Foil & $5.6\times10^{10}$ & $7.8\times10^{8}$ & $1.8\times 10^{-3}$ \\
    \hline&&&\\[-1em]
         µ-bar ($w$~=~2~µm)& $3.8\times10^{10}$  &$1.1\times10^{9}$&$7.4\times 10^{-3}$  \\
    \hline
    \end{tabular}
    \caption{Comparison of the flux, number and energy conversion efficiency for protons with energy above 1~MeV, emitted from foil vs. µ-bar targets.}
    \label{tab:my_label}
\end{table}

\subsubsection*{Cascaded acceleration}
\begin{figure}[h]%
\centering  
\includegraphics[width=0.65\textwidth]{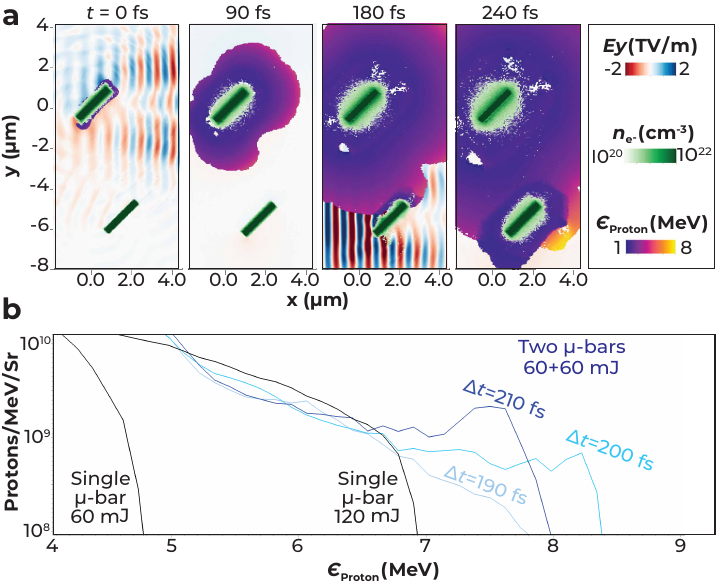}
\caption{3D PIC simulation of a cascaded proton accelerator based on the irradiation of two µ-bars.
 Each µ-bar was irradiated with a 25~fs long, 60 mJ laser pulse at a 45$^\circ$ angle of incidence. 
 The second pulse follows the first one with a $\Delta t$~=~200~fs delay.
(a) Snapshots of the transverse component of the electric field ($E_y$) are shown in a red-to-blue color scale, with the electron density overlaid in green
and the space-averaged energy of protons in a color scale.
At $t$~=~240~fs, protons that were emitted from the first µ-bar are observed to be further accelerated by the sheath of the second µ-bar  to energies above 8 MeV. 
(b) The resulting differential proton energy spectra from the simulation shown in (a), 
compared to simulations with shorter 
($\Delta t$ = 190) or longer ($\Delta t$ = 210) delays between the two pulses.
The spectra are also compared with those resulting from the irradiation of a single µ-bar with a   
25~fs long laser pulse of 60 or 120 mJ.}
\label{fig:two-bars}
\end{figure}
The use of micrometric formations as targets for TNSA of ions suggests the possibility of 
cascaded ion acceleration by sequential irradiation of multiple targets.
This method has been attempted using solid foils \cite{pfotenhauer2010cascaded,wang2018multi}, but because the numerical aperture in this irradiation scenario is on the order of unity,
 the separation distance between the two targets must be kept close to their lateral dimension
 to allow a second laser pulse to fit between the targets. 
 Indeed, the separation between the foil targets in those experiments was on the order of millimeters, resulting in a dispersed proton bunch that is much larger than the sheath of the second target, thus making the secondary acceleration inefficient.

We demonstrate numerically the potential of a cascaded proton accelerator composed of micrometric targets in Fig.~\ref{fig:two-bars}. 
Two $d$ = 0.2 µm µ-bars are positioned parallel to their $w$ = 2 µm sides, with a separation of 6 µm between them. 
The targets are irradiated at a 45$^\circ$ angle of incidence with two laser pulses,  having the same parameters as in the simulations presented in Fig.~\ref{fig:pic}, with the laser energy distributed evenly between them. 
The second pulse is delayed by $\Delta t$~=~200~fs with respect to the first, so an electron sheath forms around the second µ-bar at the time of arrival of $\sim$4 MeV protons emitted from the first target.
The position of the second µ-bar is chosen at the minimal separation in which the intensity of the first pulse would not induce a sheath around it prematurely, taking advantage of the increased divergence of the proton beam discussed above.
Fig.~\ref{fig:two-bars}a shows four snapshot of the transverse component of the laser field (red-to-blue), electron density (green), and the space-averaged proton energy (color scale). 
The final snapshot demonstrates how the sheath around the second target further accelerates those protons to energies of over 8 MeV.
The resulting differential energy spectra of the emitted protons is shown in Fig.~\ref{fig:two-bars}b. 
It is also compared to two identical simulations in which the second pulse arrive too early ($\Delta t$~=~190~fs) or too late ($\Delta t$~=~210~fs),
resulting in lower proton cutoff energies. 
  
In comparison to the double-foil target experiments discussed above, we may consider for example a beam of protons emitted from the first target with energies close to 5 MeV, and a 1 µm thick sheath prevailing for 50~fs of effective acceleration time \cite{fuchs2006laser} around the second target.
A 6-µm separation distance will correspond to effective acceleration of
$5\pm0.64$~MeV protons, while a 1-mm separation distance
will only accelerate protons arriving within $5\pm0.004$~MeV.
Target assemblies consisting of multiple µ-bars having set separation distances on the µm scale, can be realized using photolithography-based micromachining  \cite{gershuni2019gatling}.
Realization of two foci that are a few µm apart, with a controllable relative delay,
can be achieved by
spatially dividing the beam into two parts before the final focusing optics,
with one part having an adjustable relative angle and front-facing position \cite{aurand2016manipulation}.
Such an optical manipulator is simple and compact, but it comes at the cost of reduced intensity due to the inherent astigmatic focusing of one of the beams.
Alternatively, an astigmatic system can be realized by expanding the beam to a sufficiently large diameter such that different spatial parts of the focusing beam can be steered with the desired relative delay, while keeping the fluence below the optical damage threshold of the optics.

For a given µ-bars separation distance and $\Delta t$ chosen under the considerations above, the energy gain of protons injected at time $t_{in}$ with respect to $\Delta t$ to the rear sheath field of the second µ-bar, can be evaluated in the isothermal plasma expansion approximation \cite{mora2003plasma}:
$m_p v \frac{dv}{dx} = eE(x, t_{in}) \approx 2 k_B T_e/(x + c_s t_{in} + \sqrt{2} \lambda_D)$.
Here $m_p$ is the proton's mass, $v$ and $x$ are its velocity and position with respect to the rear surface, $\lambda_D$ is the Debye length, and $c_s$ is the ion sound speed.
For simplicity, we assumed above that the sheath field is quasi-static during the proton’s interaction (i.e., “frozen” at time \( t_{\text{in}} \)).
Separation of variables ($x$ and $v$) and integration of both sides of the above expression yields:
$\Delta \mathcal{E} = \frac{1}{2} m_p (v_{\infty}^2 - v_{in}^2) = 2 k_B T_e \ln \left( \frac{2 c_s t_{\text{in}} + \sqrt{2} \lambda_D}{c_s t_{\text{in}} + \sqrt{2} \lambda_D} \right)$,
where we assumed that the sheath field extends only up to the ion front at $x \approx c_s t_{in}$.
This implies an energy gain linear in $t_{in}$ at early times ($c_s t_{in} \ll \lambda_D$),
$\Delta \mathcal{E} \approx \sqrt{2}  k_B T_e c_s t_{in}/\lambda_D$,
and asymptotic at late times ($c_s t_{in} \gg \lambda_D$), $\Delta \mathcal{E} \approx  2  \ln(2) k_B T_e $.

\subsubsection*{Practical aspects}
A few practical aspects in the irradiation of µ-bar targets need consideration. 
First, unlike planar targets which are transversely larger than the focus dimensions, µ-bars are more susceptible to intensity drops resulting from the pointing instability of the laser system. 
Second, while proton production targets in the form of
massive rotating disks \cite{hou2009mev},
spooled tape \cite{noaman2017statistical},
and jets of
liquids, gases, and molecular clusters \cite{morrison2018mev,treffert2022high,willingale2006collimated, fukuda2009energy}
demonstrated target replenishment with Hz -- kHz rates,
systems designed to deliver micromachined targets mounted on either whole wafers \cite{gershuni2019gatling}
or small chips \cite{cohen2024accumulated} operate at sub-Hz rates only and their ability to position targets with submicrometer accuracy in the transverse direction
has yet to be demonstrated.
The results presented here should motivate the development of liquid jets producing replenishing targets that are bound and stable in the transverse direction within the µm scale.

Targets having a large surface covered with micrometric formations \cite{Zigler2013,Floquet2013,curtis2018micro,ceccotti2013evidence,sgattoni2015laser,blanco2017table,Khaghani2017,Frontiers2023} are naturally more resilient to the laser's miss-pointing making them simpler and faster to align. 
This advantage is shared also with many other target designs which manipulate laser absorption to enhance TNSA \cite{Sgattoni2012,prencipe2021efficient,Jiang2018}.
However, the stochastic nature of the overlap between their micrometric features and the laser field results in an inherent shot-to-shot variability of the accelerated ions. 
Beyond the increase in the proton energies and the reduced source-size reported here, 
single-formation targets like µ-bars and levitated spheres \cite{Ostermayr2016spheres} unveil the interaction dynamics in a straightforward manner by allowing parametric scans over their geometrical features. 

Finally, 
understanding the sensitivity of the acceleration to the intensity of precursor light is paramount to evaluating the scalability of our findings to larger laser systems. 
This future study will be conducted through time-resolved plasma interferometry of the target's pre-expansion.

In summary, we discovered that the interaction of an intense laser pulse with an object whose dimensions are transversely immersed in the focal volume and thinner than half the laser wavelength results in enhanced TNSA.
The smaller target dimensions provide a small virtual source size and very low emittance.
By irradiating 2~µm wide, 0.2~µm thick gold bar targets with 120~mJ of laser energy, 
we accelerated protons to over 6~MeV, which is three times higher than those obtained with a conventional planar foil target.
We note that for the acceleration of ions heavier than $H^+$, other methods such as the Coloumb explosion of molecular clusters \cite{fukuda2009energy} were able to generate even higher ion energies using the same laser pulse energy. 

This irradiation scheme makes possible cascaded acceleration by irradiation of multiple targets with micrometric spacing, which could provide even higher proton energies and optical means to control their spectrum.

\section*{ Methods}
\noindent\textbf{Target fabrication}

The targets were free-standing Au bars suspended over rectangular openings in a 250 µm thick Si wafer support. 
The fabrication process started with a Si wafer pre-coated on its front with a 200-nm thick layer of high-stress Si$_3$N$_4$. 
The back side of the wafer was spin-coated with layers of resist (MicroChem SF9) and photoresist (MicroChem AZ-1518),
on which 3.0 mm $\times$ 0.4 mm rectangular gaps were photo-lithographed.
The Si was then etched in a 30\% KOH solution at 90$^{\circ}$c.
The process stopped spontaneously when the inner surface of the front side Si$_3$N$_4$ was exposed.
Next, the Si$_3$N$_4$ side of the wafer was spin-coated with layers of the same resist and photoresist. 
1.7--10.5 µm wide rectangular openings, which would form the micro-bars,
were photo-lithographed over the gaps.
The wafer was coated with a 10-nm thick Ti adhesion layer and a 190-nm thick layer of Au. 
The Si$_3$N$_4$ around the bars was removed by reactive ion etching and immersed in Acetone. 
Finally, the remaining Si$_3$N$_4$ layer below the Au bars was removed by dry-etching.  
Illustrations of parts of the process are given in Ref.~\cite{gershuni2019gatling}. \\

\noindent\textbf{PIC simulation}

We used the fully relativistic EPOCH PIC code \cite{arber2015contemporary} to carry out the simulations. 
In these simulations, $d$ = 0.2 µm thick µ-bars of various widths were irradiated with p-polarized 800-nm wavelength laser pulses having a 30 fs (FWHM) wide Gaussian temporal profile and 120 mJ of energy.
The laser pulses were focused to a spot size of 3.5 µm (FWHM), yielding a normalized laser amplitude  of $a_0$~=~4.6.
The 3D simulation space was defined as a (32 µm)$_x\times$(20 µm)$_y\times$(24 µm)$_z$ box divided into (1000)$_x\times$(1000)$_y\times$(150)$_z$  computational mesh cells.
We conducted one computationally heavy simulation with a high resolution of (8000)$_x\times$(3000)$_y\times$(300)$_z$ cells to verify the consistency of the results.

The focused laser intensity is sufficient to ionize the target to charge states of up to $\sim$Au$^{40+}$ \cite{kawahito2020ionization}.
In the simulations presented in this paper, the bulk of the targets was representative of Au$^{20+}$ ions and electrons with densities of 30 and 20$\times$30 times that of the critical plasma density respectively.
The targets were surrounded on all sides by an exponential density gradient with a scale length of $\lambda$/60. 
This evaluation relies on findings of Ref.~\cite{cantono2021laser} which studied proton emission from ultrathin Au foils irradiated  
with similar laser intensity and temporal contrast. Using a simple sonic-expansion  model they estimated a scale length in the range of 20--70~nm, which yielded minimal modifications to the resulting proton spectrum.
Rerunning the simulations with initial ion charge states in the range of  Au$^{4+}$--Au$^{40+}$, 
and with a pre-plasma scale length in the range of  $\lambda$/80 --  $\lambda$/40, resulted in an overall shift of the proton energies by a factor of 0.88--1.24,
but the dependence on $w$ remained unchanged.
Following the results of this sensitivity test, dynamical treatment of the ionization was not simulated to reduce computational load. 

An external contaminate layer composed of H$^+$, C$^{4+}$, and O$^{4+}$ ions in equal parts was set with a uniform density 30 times that of the critical plasma density over a thickness of 0.1 µm \cite{Kluge2010mass-limited-foils}. 
The distribution of the composition of the target along the long dimension of the µ-bar was uniform over the range of $|z|<$ 5.5 µm.\\

\noindent\textbf{Experimental setup}

\begin{figure*}[h]%
\centering  
\includegraphics[width=0.48\textwidth]{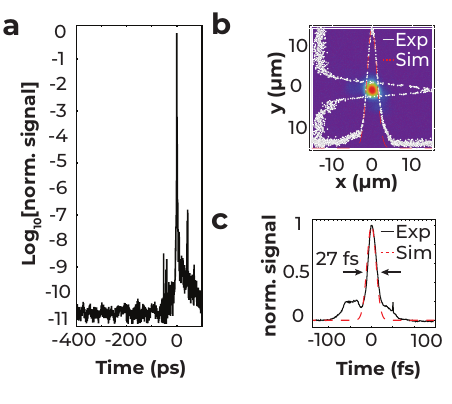}
\caption{(a) The laser's temporal profile measured with a sequoia third-order autocorrelator (Sequoia HD, Amplitude Tech.).
 (b) The measured (white) and simulated (red) profiles of the focal intensity,
 overlayed over an x100 magnification image of the focus taken at low-power, with the same scale as the $r$ axis.
 (c) The measured (black) and simulated (red) profiles of the laser pulse duration. The measured trace was obtained using a 2nd order autocorrelator.}
\label{fig:pulse-params}
\end{figure*}

27-fs long laser pulses of central wavelength $\lambda$~=~800~nm, with energies of 120 mJ (on-target) and pulse contrast better than $10^{11}$ before $t = -60$~ps \cite{PhysRevResearch.3.L032059},
 that are polarized along the width ($w$) of the bar,
were focused using an f/2.5 off-axis parabolic mirror unto $d$ = 0.2 µm thick and $w$~=~2~--~10~µm wide µ-bars. 
70\% of the laser energy was measured to be contained within a circle of 3.5 µm diameter, corresponding to a normalized laser amplitude of $a_0$~=~4.6.
The laser pointing stability was measured to be 0.43 µm (RMS).
See 2nd and 3rd order autocorrelation trace measurements, and a measurement of the focal spot in low power in Fig.~\ref{fig:pulse-params}.

\begin{figure*}[h]%
\centering  
\includegraphics[width=0.7\textwidth]{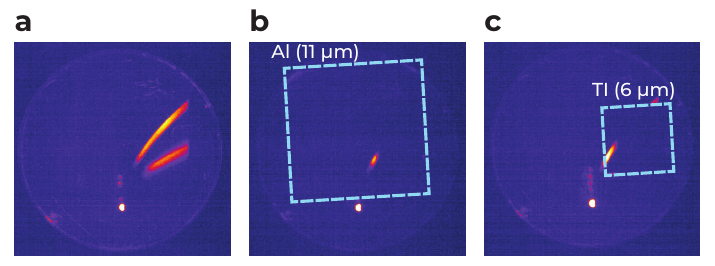}
\caption{Position-to-proton-energy calibration. 
Raw TPIS traces for irradiation experiments under identical conditions, for cases where the scintillator was  
(a) uncovered,
(b) covered with a 11 µm thick Al foil, and
(c) covered with a 6 µm thick Ti foil.
The dashed frame indicates the position of the filter.}
\label{fig:TP_callibration}
\end{figure*}

We recorded ion spectra using a TPIS 
with a similar design to that of Morrison et al. \cite{morrison2011design},
operating with an electrode voltage difference of 2 kV.
The spectrometer aperture was set to accept ions arriving at a solid angle of 3.56 µsr around the laser propagation direction.
A charge-coupled device imaged a CsI(Tl) scintillator \cite{Pappalardo2010} positioned at the end of the spectrometer. 
This type of scintillator is very suitable for laser-based particle acceleration experiments,
as it is very bright both for MeV-level electrons \cite{cohen2024undepleted} and ions \cite{gershuni2019gatling}, and its peak emission is in the visible spectrum (540 nm).
Calibration of the position along the parabolic trace of the protons to absolute energy was obtained by taking shots with parts of the scintillator covered by foil filters of known thickness and composition.
Fig.~\ref{fig:TP_callibration}, presents raw TPIS traces for irradiation experiments under identical conditions, for cases in which the scintillator was  
(a) uncovered,
(b) covered with a 11 µm thick Al foil, and
(c) covered with a 6 µm thick Ti foil.
The edges of the filter foils are indicated by a dashed  frame.
Using calculated punch-through energy values of protons for each of the filters \cite{srim}, 0.85 MeV for 11 µm thick Al and 0.65 MeV for 6 µm thick Ti,  each measurement provides one absolute energy calibration point.
We calibrated the conversion of the scintillation signal to an absolute proton dose value by recording spectrometer traces using image plates \cite{gershuni2019gatling,manvcic2008absolute}.

\section*{Data availability}
The authors declare that all data supporting the findings of this study are
available within the paper.

\bibliography{ubars-bibliography}


\begin{thebibliography}{77}
\ifx \bisbn   \undefined \def \bisbn  #1{ISBN #1}\fi
\ifx \binits  \undefined \def \binits#1{#1}\fi
\ifx \bauthor  \undefined \def \bauthor#1{#1}\fi
\ifx \batitle  \undefined \def \batitle#1{#1}\fi
\ifx \bjtitle  \undefined \def \bjtitle#1{#1}\fi
\ifx \bvolume  \undefined \def \bvolume#1{\textbf{#1}}\fi
\ifx \byear  \undefined \def \byear#1{#1}\fi
\ifx \bissue  \undefined \def \bissue#1{#1}\fi
\ifx \bfpage  \undefined \def \bfpage#1{#1}\fi
\ifx \blpage  \undefined \def \blpage #1{#1}\fi
\ifx \burl  \undefined \def \burl#1{\textsf{#1}}\fi
\ifx \doiurl  \undefined \def \doiurl#1{\url{https://doi.org/#1}}\fi
\ifx \betal  \undefined \def \betal{\textit{et al.}}\fi
\ifx \binstitute  \undefined \def \binstitute#1{#1}\fi
\ifx \binstitutionaled  \undefined \def \binstitutionaled#1{#1}\fi
\ifx \bctitle  \undefined \def \bctitle#1{#1}\fi
\ifx \beditor  \undefined \def \beditor#1{#1}\fi
\ifx \bpublisher  \undefined \def \bpublisher#1{#1}\fi
\ifx \bbtitle  \undefined \def \bbtitle#1{#1}\fi
\ifx \bedition  \undefined \def \bedition#1{#1}\fi
\ifx \bseriesno  \undefined \def \bseriesno#1{#1}\fi
\ifx \blocation  \undefined \def \blocation#1{#1}\fi
\ifx \bsertitle  \undefined \def \bsertitle#1{#1}\fi
\ifx \bsnm \undefined \def \bsnm#1{#1}\fi
\ifx \bsuffix \undefined \def \bsuffix#1{#1}\fi
\ifx \bparticle \undefined \def \bparticle#1{#1}\fi
\ifx \barticle \undefined \def \barticle#1{#1}\fi
\bibcommenthead
\ifx \bconfdate \undefined \def \bconfdate #1{#1}\fi
\ifx \botherref \undefined \def \botherref #1{#1}\fi
\ifx \url \undefined \def \url#1{\textsf{#1}}\fi
\ifx \bchapter \undefined \def \bchapter#1{#1}\fi
\ifx \bbook \undefined \def \bbook#1{#1}\fi
\ifx \bcomment \undefined \def \bcomment#1{#1}\fi
\ifx \oauthor \undefined \def \oauthor#1{#1}\fi
\ifx \citeauthoryear \undefined \def \citeauthoryear#1{#1}\fi
\ifx \endbibitem  \undefined \def \endbibitem {}\fi
\ifx \bconflocation  \undefined \def \bconflocation#1{#1}\fi
\ifx \arxivurl  \undefined \def \arxivurl#1{\textsf{#1}}\fi
\csname PreBibitemsHook\endcsname

\bibitem[\protect\citeauthoryear{Schaeffer et~al.}{2023}]{RevModPhys.95.045007}
\begin{barticle}
\bauthor{\bsnm{Schaeffer}, \binits{D.B.}},
\bauthor{\bsnm{Bott}, \binits{A.F.A.}},
\bauthor{\bsnm{Borghesi}, \binits{M.}},
\bauthor{\bsnm{Flippo}, \binits{K.A.}},
\bauthor{\bsnm{Fox}, \binits{W.}},
\bauthor{\bsnm{Fuchs}, \binits{J.}},
\bauthor{\bsnm{Li}, \binits{C.}},
\bauthor{\bsnm{S\'eguin}, \binits{F.H.}},
\bauthor{\bsnm{Park}, \binits{H.-S.}},
\bauthor{\bsnm{Tzeferacos}, \binits{P.}},
\bauthor{\bsnm{Willingale}, \binits{L.}}:
\batitle{Proton imaging of high-energy-density laboratory plasmas}.
\bjtitle{Rev. Mod. Phys.}
\bvolume{95},
\bfpage{045007}
(\byear{2023})
\doiurl{10.1103/RevModPhys.95.045007}
\end{barticle}
\endbibitem

\bibitem[\protect\citeauthoryear{Fern{\'a}ndez et~al.}{2014}]{fernandez2014fast}
\begin{barticle}
\bauthor{\bsnm{Fern{\'a}ndez}, \binits{J.}},
\bauthor{\bsnm{Albright}, \binits{B.}},
\bauthor{\bsnm{Beg}, \binits{F.N.}},
\bauthor{\bsnm{Foord}, \binits{M.E.}},
\bauthor{\bsnm{Hegelich}, \binits{B.M.}},
\bauthor{\bsnm{Honrubia}, \binits{J.}},
\bauthor{\bsnm{Roth}, \binits{M.}},
\bauthor{\bsnm{Stephens}, \binits{R.B.}},
\bauthor{\bsnm{Yin}, \binits{L.}}:
\batitle{Fast ignition with laser-driven proton and ion beams}.
\bjtitle{Nuclear fusion}
\bvolume{54}(\bissue{5}),
\bfpage{054006}
(\byear{2014})
\end{barticle}
\endbibitem

\bibitem[\protect\citeauthoryear{Roth et~al.}{2013}]{roth2013bright}
\begin{barticle}
\bauthor{\bsnm{Roth}, \binits{M.}},
\bauthor{\bsnm{Jung}, \binits{D.}},
\bauthor{\bsnm{Falk}, \binits{K.}},
\bauthor{\bsnm{Guler}, \binits{N.}},
\bauthor{\bsnm{Deppert}, \binits{O.}},
\bauthor{\bsnm{Devlin}, \binits{M.}},
\bauthor{\bsnm{Favalli}, \binits{A.}},
\bauthor{\bsnm{Fernandez}, \binits{J.}},
\bauthor{\bsnm{Gautier}, \binits{D.}},
\bauthor{\bsnm{Geissel}, \binits{M.}}, \betal:
\batitle{Bright laser-driven neutron source based on the relativistic transparency of solids}.
\bjtitle{Physical review letters}
\bvolume{110}(\bissue{4}),
\bfpage{044802}
(\byear{2013})
\end{barticle}
\endbibitem

\bibitem[\protect\citeauthoryear{Kishon et~al.}{2019}]{kishon2019laser}
\begin{barticle}
\bauthor{\bsnm{Kishon}, \binits{I.}},
\bauthor{\bsnm{Kleinschmidt}, \binits{A.}},
\bauthor{\bsnm{Schanz}, \binits{V.}},
\bauthor{\bsnm{Tebartz}, \binits{A.}},
\bauthor{\bsnm{Noam}, \binits{O.}},
\bauthor{\bsnm{Fernandez}, \binits{J.}},
\bauthor{\bsnm{Gautier}, \binits{D.}},
\bauthor{\bsnm{Johnson}, \binits{R.}},
\bauthor{\bsnm{Shimada}, \binits{T.}},
\bauthor{\bsnm{Wurden}, \binits{G.}}, \betal:
\batitle{Laser based neutron spectroscopy}.
\bjtitle{Nuclear Instruments and Methods in Physics Research Section A: Accelerators, Spectrometers, Detectors and Associated Equipment}
\bvolume{932},
\bfpage{27}--\blpage{30}
(\byear{2019})
\end{barticle}
\endbibitem

\bibitem[\protect\citeauthoryear{Ledingham et~al.}{2014}]{ledingham2014towards}
\begin{barticle}
\bauthor{\bsnm{Ledingham}, \binits{K.W.}},
\bauthor{\bsnm{Bolton}, \binits{P.R.}},
\bauthor{\bsnm{Shikazono}, \binits{N.}},
\bauthor{\bsnm{Ma}, \binits{C.-M.C.}}:
\batitle{Towards laser driven hadron cancer radiotherapy: A review of progress}.
\bjtitle{Applied Sciences}
\bvolume{4}(\bissue{3}),
\bfpage{402}--\blpage{443}
(\byear{2014})
\end{barticle}
\endbibitem

\bibitem[\protect\citeauthoryear{Kumada}{2020}]{kumada2020accelerator}
\begin{botherref}
\oauthor{\bsnm{Kumada}, \binits{H.}}:
Accelerator systems for proton radiotherapy.
Proton Beam Radiotherapy: Physics and Biology,
85--96
(2020)
\end{botherref}
\endbibitem

\bibitem[\protect\citeauthoryear{Passoni et~al.}{2010}]{passoni2010target}
\begin{barticle}
\bauthor{\bsnm{Passoni}, \binits{M.}},
\bauthor{\bsnm{Bertagna}, \binits{L.}},
\bauthor{\bsnm{Zani}, \binits{A.}}:
\batitle{Target normal sheath acceleration: theory, comparison with experiments and future perspectives}.
\bjtitle{New Journal of Physics}
\bvolume{12}(\bissue{4}),
\bfpage{045012}
(\byear{2010})
\end{barticle}
\endbibitem

\bibitem[\protect\citeauthoryear{Zimmer et~al.}{2021}]{zimmer2021analysis}
\begin{barticle}
\bauthor{\bsnm{Zimmer}, \binits{M.}},
\bauthor{\bsnm{Scheuren}, \binits{S.}},
\bauthor{\bsnm{Ebert}, \binits{T.}},
\bauthor{\bsnm{Schaumann}, \binits{G.}},
\bauthor{\bsnm{Schmitz}, \binits{B.}},
\bauthor{\bsnm{Hornung}, \binits{J.}},
\bauthor{\bsnm{Bagnoud}, \binits{V.}},
\bauthor{\bsnm{R{\"o}del}, \binits{C.}},
\bauthor{\bsnm{Roth}, \binits{M.}}:
\batitle{Analysis of laser-proton acceleration experiments for development of empirical scaling laws}.
\bjtitle{Physical Review E}
\bvolume{104}(\bissue{4}),
\bfpage{045210}
(\byear{2021})
\end{barticle}
\endbibitem

\bibitem[\protect\citeauthoryear{Ziegler et~al.}{2024}]{ziegler2024laser}
\begin{botherref}
\oauthor{\bsnm{Ziegler}, \binits{T.}},
\oauthor{\bsnm{G{\"o}thel}, \binits{I.}},
\oauthor{\bsnm{Assenbaum}, \binits{S.}},
\oauthor{\bsnm{Bernert}, \binits{C.}},
\oauthor{\bsnm{Brack}, \binits{F.-E.}},
\oauthor{\bsnm{Cowan}, \binits{T.E.}},
\oauthor{\bsnm{Dover}, \binits{N.P.}},
\oauthor{\bsnm{Gaus}, \binits{L.}},
\oauthor{\bsnm{Kluge}, \binits{T.}},
\oauthor{\bsnm{Kraft}, \binits{S.}}, et al.:
Laser-driven high-energy proton beams from cascaded acceleration regimes.
Nature Physics,
1--6
(2024)
\end{botherref}
\endbibitem

\bibitem[\protect\citeauthoryear{Gershuni et~al.}{2019}]{gershuni2019gatling}
\begin{barticle}
\bauthor{\bsnm{Gershuni}, \binits{Y.}},
\bauthor{\bsnm{Roitman}, \binits{D.}},
\bauthor{\bsnm{Cohen}, \binits{I.}},
\bauthor{\bsnm{Porat}, \binits{E.}},
\bauthor{\bsnm{Danan}, \binits{Y.}},
\bauthor{\bsnm{Elkind}, \binits{M.}},
\bauthor{\bsnm{Levanon}, \binits{A.}},
\bauthor{\bsnm{Louzon}, \binits{R.}},
\bauthor{\bsnm{Reichenberg}, \binits{D.}},
\bauthor{\bsnm{Tsabary}, \binits{A.}}, \betal:
\batitle{A gatling-gun target delivery system for high-intensity laser irradiation experiments}.
\bjtitle{Nuclear Instruments and Methods in Physics Research Section A: Accelerators, Spectrometers, Detectors and Associated Equipment}
\bvolume{934},
\bfpage{58}--\blpage{62}
(\byear{2019})
\end{barticle}
\endbibitem

\bibitem[\protect\citeauthoryear{Maffini et~al.}{2023}]{maffini2023laser}
\begin{barticle}
\bauthor{\bsnm{Maffini}, \binits{A.}},
\bauthor{\bsnm{Mirani}, \binits{F.}},
\bauthor{\bsnm{Giovannelli}, \binits{A.C.}},
\bauthor{\bsnm{Formenti}, \binits{A.}},
\bauthor{\bsnm{Passoni}, \binits{M.}}:
\batitle{Laser-driven production with advanced targets of copper-64 for medical applications}.
\bjtitle{Frontiers in Physics}
\bvolume{11},
\bfpage{1223023}
(\byear{2023})
\doiurl{10.3389/fphy.2023.1223023}
\end{barticle}
\endbibitem

\bibitem[\protect\citeauthoryear{Molloy et~al.}{2025}]{PhysRevResearch.7.013230}
\begin{barticle}
\bauthor{\bsnm{Molloy}, \binits{D.P.}},
\bauthor{\bsnm{Orecchia}, \binits{D.}},
\bauthor{\bsnm{Tosca}, \binits{M.}},
\bauthor{\bsnm{Milani}, \binits{A.}},
\bauthor{\bsnm{Valt}, \binits{M.}},
\bauthor{\bsnm{McNamee}, \binits{A.}},
\bauthor{\bsnm{Fitzpatrick}, \binits{C.R.J.}}, \betal:
\batitle{Alpha particle production from novel targets via laser-driven proton-boron fusion}.
\bjtitle{Phys. Rev. Research}
\bvolume{7},
\bfpage{013230}
(\byear{2025})
\doiurl{10.1103/PhysRevResearch.7.013230}
\end{barticle}
\endbibitem

\bibitem[\protect\citeauthoryear{Batani et~al.}{2025}]{batani2025generation}
\begin{barticle}
\bauthor{\bsnm{Batani}, \binits{K.L.}},
\bauthor{\bsnm{Rodrigues}, \binits{M.R.}},
\bauthor{\bsnm{Bonasera}, \binits{A.}},
\bauthor{\bsnm{Cipriani}, \binits{M.}},
\bauthor{\bsnm{Consoli}, \binits{F.}},
\bauthor{\bsnm{Filippi}, \binits{F.}},
\bauthor{\bsnm{Scisci{\`o}}, \binits{M.M.}},
\bauthor{\bsnm{Giuffrida}, \binits{L.}},
\bauthor{\bsnm{Kantarelou}, \binits{V.}},
\bauthor{\bsnm{Stancek}, \binits{S.}}, \betal:
\batitle{Generation of radioisotopes for medical applications using high-repetition, high-intensity lasers}.
\bjtitle{High Power Laser Science and Engineering}
\bvolume{13},
\bfpage{11}
(\byear{2025})
\end{barticle}
\endbibitem

\bibitem[\protect\citeauthoryear{Mirani et~al.}{2023}]{PhysRevApplied.19.044020}
\begin{barticle}
\bauthor{\bsnm{Mirani}, \binits{F.}},
\bauthor{\bsnm{Maffini}, \binits{A.}},
\bauthor{\bsnm{Passoni}, \binits{M.}}:
\batitle{Laser-driven neutron generation with near-critical targets and application to materials characterization}.
\bjtitle{Phys. Rev. Applied}
\bvolume{19},
\bfpage{044020}
(\byear{2023})
\doiurl{10.1103/PhysRevApplied.19.044020}
\end{barticle}
\endbibitem

\bibitem[\protect\citeauthoryear{Mirani et~al.}{2021}]{mirani2021integrated}
\begin{barticle}
\bauthor{\bsnm{Mirani}, \binits{F.}},
\bauthor{\bsnm{Maffini}, \binits{A.}},
\bauthor{\bsnm{Casamichiela}, \binits{F.}},
\bauthor{\bsnm{Pazzaglia}, \binits{A.}},
\bauthor{\bsnm{Formenti}, \binits{A.}},
\bauthor{\bsnm{Dellasega}, \binits{D.}}, \betal:
\batitle{Integrated quantitative pixe analysis and edx spectroscopy using a laser-driven particle source}.
\bjtitle{Science Advances}
\bvolume{7}(\bissue{3}),
\bfpage{8660}
(\byear{2021})
\doiurl{10.1126/sciadv.abc8660}
\end{barticle}
\endbibitem

\bibitem[\protect\citeauthoryear{Sgattoni et~al.}{2012}]{Sgattoni2012}
\begin{barticle}
\bauthor{\bsnm{Sgattoni}, \binits{A.}},
\bauthor{\bsnm{Londrillo}, \binits{P.}},
\bauthor{\bsnm{Macchi}, \binits{A.}},
\bauthor{\bsnm{Passoni}, \binits{M.}}:
\batitle{Laser ion acceleration using a solid target coupled with a low-density layer}.
\bjtitle{Physical Review E}
\bvolume{85},
\bfpage{036405}
(\byear{2012})
\end{barticle}
\endbibitem

\bibitem[\protect\citeauthoryear{Prencipe et~al.}{2021}]{prencipe2021efficient}
\begin{barticle}
\bauthor{\bsnm{Prencipe}, \binits{I.}},
\bauthor{\bsnm{Metzkes-Ng}, \binits{J.}},
\bauthor{\bsnm{Pazzaglia}, \binits{A.}},
\bauthor{\bsnm{Bernert}, \binits{C.}},
\bauthor{\bsnm{Dellasega}, \binits{D.}},
\bauthor{\bsnm{Fedeli}, \binits{L.}},
\bauthor{\bsnm{Formenti}, \binits{A.}},
\bauthor{\bsnm{Garten}, \binits{M.}},
\bauthor{\bsnm{Kluge}, \binits{T.}},
\bauthor{\bsnm{Kraft}, \binits{S.}}, \betal:
\batitle{Efficient laser-driven proton and bremsstrahlung generation from cluster-assembled foam targets}.
\bjtitle{New Journal of Physics}
\bvolume{23}(\bissue{9}),
\bfpage{093015}
(\byear{2021})
\end{barticle}
\endbibitem

\bibitem[\protect\citeauthoryear{Jiang et~al.}{2018}]{Jiang2018}
\begin{barticle}
\bauthor{\bsnm{Jiang}, \binits{S.Z.}},
\bauthor{\bsnm{Yu}, \binits{W.}},
\bauthor{\bsnm{Wang}, \binits{W.H.}},
\bauthor{\bsnm{Cao}, \binits{L.}},
\bauthor{\bsnm{Li}, \binits{Y.T.}},
\bauthor{\bsnm{Qiao}, \binits{B.}},
\bauthor{\bsnm{Ruan}, \binits{S.}},
\bauthor{\bsnm{He}, \binits{X.T.}}:
\batitle{Target normal sheath acceleration of protons using triple-layer target}.
\bjtitle{Physical Review Accelerators and Beams}
\bvolume{21},
\bfpage{051303}
(\byear{2018})
\doiurl{10.1103/PhysRevAccelBeams.21.051303}
\end{barticle}
\endbibitem

\bibitem[\protect\citeauthoryear{Ceccotti et~al.}{2013}]{ceccotti2013evidence}
\begin{barticle}
\bauthor{\bsnm{Ceccotti}, \binits{T.}},
\bauthor{\bsnm{Floquet}, \binits{V.}},
\bauthor{\bsnm{Sgattoni}, \binits{A.}},
\bauthor{\bsnm{Bigongiari}, \binits{A.}},
\bauthor{\bsnm{Klimo}, \binits{O.}},
\bauthor{\bsnm{Raynaud}, \binits{M.}},
\bauthor{\bsnm{Riconda}, \binits{C.}},
\bauthor{\bsnm{Heron}, \binits{A.}},
\bauthor{\bsnm{Baffigi}, \binits{F.}},
\bauthor{\bsnm{Labate}, \binits{L.}}, \betal:
\batitle{Evidence of resonant surface-wave excitation in the relativistic regime through measurements of proton acceleration from grating targets}.
\bjtitle{Physical review letters}
\bvolume{111}(\bissue{18}),
\bfpage{185001}
(\byear{2013})
\end{barticle}
\endbibitem

\bibitem[\protect\citeauthoryear{Sgattoni et~al.}{2015}]{sgattoni2015laser}
\begin{barticle}
\bauthor{\bsnm{Sgattoni}, \binits{A.}},
\bauthor{\bsnm{Sinigardi}, \binits{S.}},
\bauthor{\bsnm{Fedeli}, \binits{L.}},
\bauthor{\bsnm{Pegoraro}, \binits{F.}},
\bauthor{\bsnm{Macchi}, \binits{A.}}:
\batitle{Laser-driven rayleigh-taylor instability: Plasmonic effects and three-dimensional structures}.
\bjtitle{Physical Review E}
\bvolume{91}(\bissue{1}),
\bfpage{013106}
(\byear{2015})
\end{barticle}
\endbibitem

\bibitem[\protect\citeauthoryear{Blanco et~al.}{2017}]{blanco2017table}
\begin{barticle}
\bauthor{\bsnm{Blanco}, \binits{M.}},
\bauthor{\bsnm{Flores-Arias}, \binits{M.T.}},
\bauthor{\bsnm{Ruiz}, \binits{C.}},
\bauthor{\bsnm{Vranic}, \binits{M.}}:
\batitle{Table-top laser-based proton acceleration in nanostructured targets}.
\bjtitle{New Journal of Physics}
\bvolume{19}(\bissue{3}),
\bfpage{033004}
(\byear{2017})
\end{barticle}
\endbibitem

\bibitem[\protect\citeauthoryear{Lezhnin and Bulanov}{2022}]{Lezhnin2022}
\begin{barticle}
\bauthor{\bsnm{Lezhnin}, \binits{K.V.}},
\bauthor{\bsnm{Bulanov}, \binits{S.V.}}:
\batitle{Laser ion acceleration from tailored solid targets with micron-scale channels}.
\bjtitle{Physical Review Research}
\bvolume{4}(\bissue{3}),
\bfpage{033248}
(\byear{2022})
\doiurl{10.1103/PhysRevResearch.4.033248}
\end{barticle}
\endbibitem

\bibitem[\protect\citeauthoryear{Zou et~al.}{2019}]{Zou2019}
\begin{barticle}
\bauthor{\bsnm{Zou}, \binits{D.B.}},
\bauthor{\bsnm{Yu}, \binits{D.Y.}},
\bauthor{\bsnm{Jiang}, \binits{X.R.}},
\bauthor{\bsnm{Yu}, \binits{M.Y.}},
\bauthor{\bsnm{Chen}, \binits{Z.Y.}},
\bauthor{\bsnm{Deng}, \binits{Z.G.}},
\bauthor{\bsnm{Yu}, \binits{T.P.}},
\bauthor{\bsnm{Yin}, \binits{Y.}},
\bauthor{\bsnm{Shao}, \binits{F.Q.}},
\bauthor{\bsnm{Zhuo}, \binits{H.B.}},
\bauthor{\bsnm{Zhou}, \binits{C.T.}},
\bauthor{\bsnm{Ruan}, \binits{S.C.}}:
\batitle{Enhancement of target normal sheath acceleration in laser multi-channel target interaction}.
\bjtitle{Physics of Plasmas}
\bvolume{26}(\bissue{12}),
\bfpage{123105}
(\byear{2019})
\doiurl{10.1063/1.5096902}
\end{barticle}
\endbibitem

\bibitem[\protect\citeauthoryear{Yang et~al.}{2018}]{Yang2018}
\begin{barticle}
\bauthor{\bsnm{Yang}, \binits{Y.C.}},
\bauthor{\bsnm{Zhou}, \binits{C.T.}},
\bauthor{\bsnm{Huang}, \binits{T.W.}},
\bauthor{\bsnm{Liu}, \binits{B.}},
\bauthor{\bsnm{He}, \binits{X.T.}}:
\batitle{Proton acceleration from laser interaction with a complex double-layer plasma target}.
\bjtitle{Physics of Plasmas}
\bvolume{25}(\bissue{12}),
\bfpage{123112}
(\byear{2018})
\doiurl{10.1063/1.5055763}
\end{barticle}
\endbibitem

\bibitem[\protect\citeauthoryear{Khaghani et~al.}{2017}]{Khaghani2017}
\begin{barticle}
\bauthor{\bsnm{Khaghani}, \binits{D.}},
\bauthor{\bsnm{Lobet}, \binits{M.}},
\bauthor{\bsnm{Borm}, \binits{B.}},
\bauthor{\bsnm{Burr}, \binits{L.}},
\bauthor{\bsnm{Gärtner}, \binits{F.}},
\bauthor{\bsnm{Gremillet}, \binits{L.}},
\bauthor{\bsnm{Movsesyan}, \binits{L.}},
\bauthor{\bsnm{Rosmej}, \binits{O.}},
\bauthor{\bsnm{Toimil-Molares}, \binits{M.E.}},
\bauthor{\bsnm{Wagner}, \binits{F.}},
\bauthor{\bsnm{Neumayer}, \binits{P.}}:
\batitle{Enhancing laser-driven proton acceleration by using micro-pillar arrays at high drive energy}.
\bjtitle{Scientific Reports}
\bvolume{7},
\bfpage{11366}
(\byear{2017})
\doiurl{10.1038/s41598-017-11589-z}
\end{barticle}
\endbibitem

\bibitem[\protect\citeauthoryear{Zhao et~al.}{2023}]{Frontiers2023}
\begin{barticle}
\bauthor{\bsnm{Zhao}, \binits{Y.}},
\bauthor{\bsnm{Li}, \binits{J.}},
\bauthor{\bsnm{Wang}, \binits{X.}},
\bauthor{\bsnm{Zhang}, \binits{L.}},
\bauthor{\bsnm{Liu}, \binits{Y.}},
\bauthor{\bsnm{Wang}, \binits{W.}},
\bauthor{\bsnm{Xu}, \binits{T.}},
\bauthor{\bsnm{Liu}, \binits{Y.}},
\bauthor{\bsnm{Liu}, \binits{J.}},
\bauthor{\bsnm{Zhang}, \binits{Y.}}, \betal:
\batitle{Enhanced laser-driven backward proton acceleration using micro-wire array targets}.
\bjtitle{Frontiers in Physics}
\bvolume{11},
\bfpage{1167927}
(\byear{2023})
\doiurl{10.3389/fphy.2023.1167927}
\end{barticle}
\endbibitem

\bibitem[\protect\citeauthoryear{Keppler et~al.}{2022}]{keppler2022intensity}
\begin{barticle}
\bauthor{\bsnm{Keppler}, \binits{S.}},
\bauthor{\bsnm{Elkina}, \binits{N.}},
\bauthor{\bsnm{Becker}, \binits{G.}},
\bauthor{\bsnm{Hein}, \binits{J.}},
\bauthor{\bsnm{Hornung}, \binits{M.}},
\bauthor{\bsnm{M{\"a}usezahl}, \binits{M.}},
\bauthor{\bsnm{R{\"o}del}, \binits{C.}},
\bauthor{\bsnm{Tamer}, \binits{I.}},
\bauthor{\bsnm{Zepf}, \binits{M.}},
\bauthor{\bsnm{Kaluza}, \binits{M.}}:
\batitle{Intensity scaling limitations of laser-driven proton acceleration in the tnsa-regime}.
\bjtitle{Physical Review Research}
\bvolume{4}(\bissue{1}),
\bfpage{013065}
(\byear{2022})
\end{barticle}
\endbibitem

\bibitem[\protect\citeauthoryear{Buffechoux et~al.}{2010}]{buffechoux2010hot}
\begin{barticle}
\bauthor{\bsnm{Buffechoux}, \binits{S.}},
\bauthor{\bsnm{Psikal}, \binits{J.}},
\bauthor{\bsnm{Nakatsutsumi}, \binits{M.}},
\bauthor{\bsnm{Romagnani}, \binits{L.}},
\bauthor{\bsnm{Andreev}, \binits{A.}},
\bauthor{\bsnm{Zeil}, \binits{K.}},
\bauthor{\bsnm{Amin}, \binits{M.}},
\bauthor{\bsnm{Antici}, \binits{P.}},
\bauthor{\bsnm{Burris-Mog}, \binits{T.}},
\bauthor{\bsnm{Compant-La-Fontaine}, \binits{A.}}, \betal:
\batitle{Hot electrons transverse refluxing in ultraintense laser-solid interactions}.
\bjtitle{Physical review letters}
\bvolume{105}(\bissue{1}),
\bfpage{015005}
(\byear{2010})
\end{barticle}
\endbibitem

\bibitem[\protect\citeauthoryear{Kraft et~al.}{2018}]{kraft2018first}
\begin{barticle}
\bauthor{\bsnm{Kraft}, \binits{S.D.}},
\bauthor{\bsnm{Obst}, \binits{L.}},
\bauthor{\bsnm{Metzkes-Ng}, \binits{J.}},
\bauthor{\bsnm{Schlenvoigt}, \binits{H.-P.}},
\bauthor{\bsnm{Zeil}, \binits{K.}},
\bauthor{\bsnm{Michaux}, \binits{S.}},
\bauthor{\bsnm{Chatain}, \binits{D.}},
\bauthor{\bsnm{Perin}, \binits{J.-P.}},
\bauthor{\bsnm{Chen}, \binits{S.N.}},
\bauthor{\bsnm{Fuchs}, \binits{J.}}, \betal:
\batitle{First demonstration of multi-mev proton acceleration from a cryogenic hydrogen ribbon target}.
\bjtitle{Plasma Physics and Controlled Fusion}
\bvolume{60}(\bissue{4}),
\bfpage{044010}
(\byear{2018})
\end{barticle}
\endbibitem

\bibitem[\protect\citeauthoryear{Zeil et~al.}{2014}]{zeil2014robust}
\begin{barticle}
\bauthor{\bsnm{Zeil}, \binits{K.}},
\bauthor{\bsnm{Metzkes}, \binits{J.}},
\bauthor{\bsnm{Kluge}, \binits{T.}},
\bauthor{\bsnm{Bussmann}, \binits{M.}},
\bauthor{\bsnm{Cowan}, \binits{T.}},
\bauthor{\bsnm{Kraft}, \binits{S.}},
\bauthor{\bsnm{Sauerbrey}, \binits{R.}},
\bauthor{\bsnm{Schmidt}, \binits{B.}},
\bauthor{\bsnm{Zier}, \binits{M.}},
\bauthor{\bsnm{Schramm}, \binits{U.}}:
\batitle{Robust energy enhancement of ultrashort pulse laser accelerated protons from reduced mass targets}.
\bjtitle{Plasma Physics and Controlled Fusion}
\bvolume{56}(\bissue{8}),
\bfpage{084004}
(\byear{2014})
\end{barticle}
\endbibitem

\bibitem[\protect\citeauthoryear{Toncian et~al.}{2011}]{toncian2011optimal}
\begin{botherref}
\oauthor{\bsnm{Toncian}, \binits{T.}},
\oauthor{\bsnm{Swantusch}, \binits{M.}},
\oauthor{\bsnm{Toncian}, \binits{M.}},
\oauthor{\bsnm{Willi}, \binits{O.}},
\oauthor{\bsnm{Andreev}, \binits{A.}},
\oauthor{\bsnm{Platonov}, \binits{K.}}:
Optimal proton acceleration from lateral limited foil sections and different laser pulse durations at relativistic intensity.
Physics of Plasmas
\textbf{18}(4)
(2011)
\end{botherref}
\endbibitem

\bibitem[\protect\citeauthoryear{Tresca et~al.}{2011}]{tresca2011controlling}
\begin{barticle}
\bauthor{\bsnm{Tresca}, \binits{O.}},
\bauthor{\bsnm{Carroll}, \binits{D.}},
\bauthor{\bsnm{Yuan}, \binits{X.}},
\bauthor{\bsnm{Aurand}, \binits{B.}},
\bauthor{\bsnm{Bagnoud}, \binits{V.}},
\bauthor{\bsnm{Brenner}, \binits{C.}},
\bauthor{\bsnm{Coury}, \binits{M.}},
\bauthor{\bsnm{Fils}, \binits{J.}},
\bauthor{\bsnm{Gray}, \binits{R.}},
\bauthor{\bsnm{K{\"u}hl}, \binits{T.}}, \betal:
\batitle{Controlling the properties of ultraintense laser--proton sources using transverse refluxing of hot electrons in shaped mass-limited targets}.
\bjtitle{Plasma physics and controlled fusion}
\bvolume{53}(\bissue{10}),
\bfpage{105008}
(\byear{2011})
\end{barticle}
\endbibitem

\bibitem[\protect\citeauthoryear{Fang et~al.}{2016}]{fang2016different}
\begin{barticle}
\bauthor{\bsnm{Fang}, \binits{Y.}},
\bauthor{\bsnm{Ge}, \binits{X.}},
\bauthor{\bsnm{Yang}, \binits{S.}},
\bauthor{\bsnm{Wei}, \binits{W.}},
\bauthor{\bsnm{Yu}, \binits{T.}},
\bauthor{\bsnm{Liu}, \binits{F.}},
\bauthor{\bsnm{Chen}, \binits{M.}},
\bauthor{\bsnm{Liu}, \binits{J.}},
\bauthor{\bsnm{Yuan}, \binits{X.}},
\bauthor{\bsnm{Sheng}, \binits{Z.}}, \betal:
\batitle{Different effects of laser contrast on proton emission from normal large foils and transverse-size-reduced targets}.
\bjtitle{Plasma Physics and Controlled Fusion}
\bvolume{58}(\bissue{7}),
\bfpage{075010}
(\byear{2016})
\end{barticle}
\endbibitem

\bibitem[\protect\citeauthoryear{Green et~al.}{2014}]{green2014high}
\begin{botherref}
\oauthor{\bsnm{Green}, \binits{J.}},
\oauthor{\bsnm{Robinson}, \binits{A.}},
\oauthor{\bsnm{Booth}, \binits{N.}},
\oauthor{\bsnm{Carroll}, \binits{D.}},
\oauthor{\bsnm{Dance}, \binits{R.}},
\oauthor{\bsnm{Gray}, \binits{R.}},
\oauthor{\bsnm{MacLellan}, \binits{D.}},
\oauthor{\bsnm{McKenna}, \binits{P.}},
\oauthor{\bsnm{Murphy}, \binits{C.}},
\oauthor{\bsnm{Rusby}, \binits{D.}}, et al.:
High efficiency proton beam generation through target thickness control in femtosecond laser-plasma interactions.
Applied Physics Letters
\textbf{104}(21)
(2014)
\end{botherref}
\endbibitem

\bibitem[\protect\citeauthoryear{Hornung et~al.}{2020}]{hornung2020enhancement}
\begin{barticle}
\bauthor{\bsnm{Hornung}, \binits{J.}},
\bauthor{\bsnm{Zobus}, \binits{Y.}},
\bauthor{\bsnm{Boller}, \binits{P.}},
\bauthor{\bsnm{Brabetz}, \binits{C.}},
\bauthor{\bsnm{Eisenbarth}, \binits{U.}},
\bauthor{\bsnm{K{\"u}hl}, \binits{T.}},
\bauthor{\bsnm{Major}, \binits{Z.}},
\bauthor{\bsnm{Ohland}, \binits{J.}},
\bauthor{\bsnm{Zepf}, \binits{M.}},
\bauthor{\bsnm{Zielbauer}, \binits{B.}}, \betal:
\batitle{Enhancement of the laser-driven proton source at phelix}.
\bjtitle{High Power Laser Science and Engineering}
\bvolume{8},
\bfpage{24}
(\byear{2020})
\end{barticle}
\endbibitem

\bibitem[\protect\citeauthoryear{Busold}{2014}]{busold2014construction}
\begin{botherref}
\oauthor{\bsnm{Busold}, \binits{S.}}:
Construction and characterization of a laser-driven proton beamline at gsi.
PhD thesis
(2014)
\end{botherref}
\endbibitem

\bibitem[\protect\citeauthoryear{Zigler et~al.}{2013}]{Zigler2013}
\begin{barticle}
\bauthor{\bsnm{Zigler}, \binits{A.}},
\bauthor{\bsnm{Eisenman}, \binits{S.}},
\bauthor{\bsnm{Botton}, \binits{M.}},
\bauthor{\bsnm{Nahum}, \binits{E.}},
\bauthor{\bsnm{Schleifer}, \binits{E.}},
\bauthor{\bsnm{Baspaly}, \binits{A.}},
\bauthor{\bsnm{Pomerantz}, \binits{I.}},
\bauthor{\bsnm{Abicht}, \binits{F.}},
\bauthor{\bsnm{Branzel}, \binits{J.}},
\bauthor{\bsnm{Priebe}, \binits{G.}},
\bauthor{\bsnm{Steinke}, \binits{S.}},
\bauthor{\bsnm{Andreev}, \binits{A.}},
\bauthor{\bsnm{Schnuerer}, \binits{M.}},
\bauthor{\bsnm{Sandner}, \binits{W.}},
\bauthor{\bsnm{Gordon}, \binits{D.}},
\bauthor{\bsnm{Sprangle}, \binits{P.}},
\bauthor{\bsnm{Ledingham}, \binits{K.W.D.}}:
\batitle{{Enhanced Proton Acceleration by an Ultrashort Laser Interaction with Structured Dynamic Plasma Targets}}.
\bjtitle{Physical review letters}
\bvolume{110}(\bissue{21}),
\bfpage{215004}
(\byear{2013})
\end{barticle}
\endbibitem

\bibitem[\protect\citeauthoryear{Floquet et~al.}{2013}]{Floquet2013}
\begin{barticle}
\bauthor{\bsnm{Floquet}, \binits{V.}},
\bauthor{\bsnm{Klimo}, \binits{O.}},
\bauthor{\bsnm{Psikal}, \binits{J.}},
\bauthor{\bsnm{Velyhan}, \binits{A.}},
\bauthor{\bsnm{Limpouch}, \binits{J.}},
\bauthor{\bsnm{Proska}, \binits{J.}},
\bauthor{\bsnm{Novotny}, \binits{F.}},
\bauthor{\bsnm{Stolcova}, \binits{L.}},
\bauthor{\bsnm{Macchi}, \binits{A.}},
\bauthor{\bsnm{Sgattoni}, \binits{A.}},
\bauthor{\bsnm{Vassura}, \binits{L.}},
\bauthor{\bsnm{Labate}, \binits{L.}},
\bauthor{\bsnm{Baffigi}, \binits{F.}},
\bauthor{\bsnm{Gizzi}, \binits{L.A.}},
\bauthor{\bsnm{Martin}, \binits{P.}},
\bauthor{\bsnm{Ceccotti}, \binits{T.}}:
\batitle{{Micro-sphere layered targets efficiency in laser driven proton acceleration}}.
\bjtitle{J. Appl. Phys}
\bvolume{114},
\bfpage{83305}
(\byear{2013})
\doiurl{10.1063/1.4819239}
\end{barticle}
\endbibitem

\bibitem[\protect\citeauthoryear{Margarone et~al.}{2012}]{margarone2012laser}
\begin{barticle}
\bauthor{\bsnm{Margarone}, \binits{D.}},
\bauthor{\bsnm{Klimo}, \binits{O.}},
\bauthor{\bsnm{Kim}, \binits{I.}},
\bauthor{\bsnm{Prokupek}, \binits{J.}},
\bauthor{\bsnm{Limpouch}, \binits{J.}},
\bauthor{\bsnm{Jeong}, \binits{T.}},
\bauthor{\bsnm{Mocek}, \binits{T.}},
\bauthor{\bsnm{Psikal}, \binits{J.}},
\bauthor{\bsnm{Kim}, \binits{H.}},
\bauthor{\bsnm{Proska}, \binits{J.}}, \betal:
\batitle{Laser-driven proton acceleration enhancement by nanostructured foils}.
\bjtitle{Physical review letters}
\bvolume{109}(\bissue{23}),
\bfpage{234801}
(\byear{2012})
\end{barticle}
\endbibitem

\bibitem[\protect\citeauthoryear{Curtis et~al.}{2018}]{curtis2018micro}
\begin{barticle}
\bauthor{\bsnm{Curtis}, \binits{A.}},
\bauthor{\bsnm{Calvi}, \binits{C.}},
\bauthor{\bsnm{Tinsley}, \binits{J.}},
\bauthor{\bsnm{Hollinger}, \binits{R.}},
\bauthor{\bsnm{Kaymak}, \binits{V.}},
\bauthor{\bsnm{Pukhov}, \binits{A.}},
\bauthor{\bsnm{Wang}, \binits{S.}},
\bauthor{\bsnm{Rockwood}, \binits{A.}},
\bauthor{\bsnm{Wang}, \binits{Y.}},
\bauthor{\bsnm{Shlyaptsev}, \binits{V.N.}}, \betal:
\batitle{Micro-scale fusion in dense relativistic nanowire array plasmas}.
\bjtitle{Nature communications}
\bvolume{9}(\bissue{1}),
\bfpage{1077}
(\byear{2018})
\end{barticle}
\endbibitem

\bibitem[\protect\citeauthoryear{Ostermayr et~al.}{2016}]{Ostermayr2016spheres}
\begin{barticle}
\bauthor{\bsnm{Ostermayr}, \binits{T.M.}},
\bauthor{\bsnm{Haffa}, \binits{D.}},
\bauthor{\bsnm{Hilz}, \binits{P.}},
\bauthor{\bsnm{Pauw}, \binits{V.}},
\bauthor{\bsnm{Allinger}, \binits{K.}},
\bauthor{\bsnm{Bamberg}, \binits{K.-U.}},
\bauthor{\bsnm{B\"ohl}, \binits{P.}},
\bauthor{\bsnm{B\"omer}, \binits{C.}},
\bauthor{\bsnm{Bolton}, \binits{P.R.}},
\bauthor{\bsnm{Deutschmann}, \binits{F.}},
\bauthor{\bsnm{Ditmire}, \binits{T.}},
\bauthor{\bsnm{Donovan}, \binits{M.E.}},
\bauthor{\bsnm{Dyer}, \binits{G.}},
\bauthor{\bsnm{Gaul}, \binits{E.}},
\bauthor{\bsnm{Gordon}, \binits{J.}},
\bauthor{\bsnm{Hegelich}, \binits{B.M.}},
\bauthor{\bsnm{Kiefer}, \binits{D.}},
\bauthor{\bsnm{Klier}, \binits{C.}},
\bauthor{\bsnm{Kreuzer}, \binits{C.}},
\bauthor{\bsnm{Martinez}, \binits{M.}},
\bauthor{\bsnm{McCary}, \binits{E.}},
\bauthor{\bsnm{Meadows}, \binits{A.R.}},
\bauthor{\bsnm{Mosch\"uring}, \binits{N.}},
\bauthor{\bsnm{R\"osch}, \binits{T.}},
\bauthor{\bsnm{Ruhl}, \binits{H.}},
\bauthor{\bsnm{Spinks}, \binits{M.}},
\bauthor{\bsnm{Wagner}, \binits{C.}},
\bauthor{\bsnm{Schreiber}, \binits{J.}}:
\batitle{Proton acceleration by irradiation of isolated spheres with an intense laser pulse}.
\bjtitle{Phys. Rev. E}
\bvolume{94},
\bfpage{033208}
(\byear{2016})
\doiurl{10.1103/PhysRevE.94.033208}
\end{barticle}
\endbibitem

\bibitem[\protect\citeauthoryear{Hilz et~al.}{2018}]{hilz2018isolated}
\begin{barticle}
\bauthor{\bsnm{Hilz}, \binits{P.}},
\bauthor{\bsnm{Ostermayr}, \binits{T.}},
\bauthor{\bsnm{Huebl}, \binits{A.}},
\bauthor{\bsnm{Bagnoud}, \binits{V.}},
\bauthor{\bsnm{Borm}, \binits{B.}},
\bauthor{\bsnm{Bussmann}, \binits{M.}},
\bauthor{\bsnm{Gallei}, \binits{M.}},
\bauthor{\bsnm{Gebhard}, \binits{J.}},
\bauthor{\bsnm{Haffa}, \binits{D.}},
\bauthor{\bsnm{Hartmann}, \binits{J.}}, \betal:
\batitle{Isolated proton bunch acceleration by a petawatt laser pulse}.
\bjtitle{Nature communications}
\bvolume{9}(\bissue{1}),
\bfpage{423}
(\byear{2018})
\end{barticle}
\endbibitem

\bibitem[\protect\citeauthoryear{Elkind et~al.}{2023}]{elkind2023intense}
\begin{barticle}
\bauthor{\bsnm{Elkind}, \binits{M.}},
\bauthor{\bsnm{Cohen}, \binits{I.}},
\bauthor{\bsnm{Blackman}, \binits{D.}},
\bauthor{\bsnm{Meir}, \binits{T.}},
\bauthor{\bsnm{Perelmutter}, \binits{L.}},
\bauthor{\bsnm{Catabi}, \binits{T.}},
\bauthor{\bsnm{Levanon}, \binits{A.}},
\bauthor{\bsnm{Glenzer}, \binits{S.H.}},
\bauthor{\bsnm{Arefiev}, \binits{A.V.}},
\bauthor{\bsnm{Pomerantz}, \binits{I.}}:
\batitle{Intense laser interaction with micro-bars}.
\bjtitle{Scientific Reports}
\bvolume{13}(\bissue{1}),
\bfpage{21345}
(\byear{2023})
\end{barticle}
\endbibitem

\bibitem[\protect\citeauthoryear{De~Andres et~al.}{2024}]{de2024unforeseen}
\begin{barticle}
\bauthor{\bsnm{De~Andres}, \binits{A.}},
\bauthor{\bsnm{Bhadoria}, \binits{S.}},
\bauthor{\bsnm{Marmolejo}, \binits{J.T.}},
\bauthor{\bsnm{Muschet}, \binits{A.}},
\bauthor{\bsnm{Fischer}, \binits{P.}},
\bauthor{\bsnm{Reza~Barzegar}, \binits{H.}},
\bauthor{\bsnm{Blackburn}, \binits{T.}},
\bauthor{\bsnm{Gonoskov}, \binits{A.}},
\bauthor{\bsnm{Hanstorp}, \binits{D.}},
\bauthor{\bsnm{Marklund}, \binits{M.}}, \betal:
\batitle{Unforeseen advantage of looser focusing in vacuum laser acceleration}.
\bjtitle{Communications Physics}
\bvolume{7}(\bissue{1}),
\bfpage{293}
(\byear{2024})
\end{barticle}
\endbibitem

\bibitem[\protect\citeauthoryear{Morrison et~al.}{2011}]{morrison2011design}
\begin{botherref}
\oauthor{\bsnm{Morrison}, \binits{J.}},
\oauthor{\bsnm{Willis}, \binits{C.}},
\oauthor{\bsnm{Freeman}, \binits{R.}},
\oauthor{\bsnm{Van~Woerkom}, \binits{L.}}:
Design of and data reduction from compact thomson parabola spectrometers.
Review of Scientific Instruments
\textbf{82}(3)
(2011)
\end{botherref}
\endbibitem

\bibitem[\protect\citeauthoryear{Arber et~al.}{2015}]{arber2015contemporary}
\begin{barticle}
\bauthor{\bsnm{Arber}, \binits{T.}},
\bauthor{\bsnm{Bennett}, \binits{K.}},
\bauthor{\bsnm{Brady}, \binits{C.}},
\bauthor{\bsnm{Lawrence-Douglas}, \binits{A.}},
\bauthor{\bsnm{Ramsay}, \binits{M.}},
\bauthor{\bsnm{Sircombe}, \binits{N.}},
\bauthor{\bsnm{Gillies}, \binits{P.}},
\bauthor{\bsnm{Evans}, \binits{R.}},
\bauthor{\bsnm{Schmitz}, \binits{H.}},
\bauthor{\bsnm{Bell}, \binits{A.}}, \betal:
\batitle{Contemporary particle-in-cell approach to laser-plasma modelling}.
\bjtitle{Plasma Physics and Controlled Fusion}
\bvolume{57}(\bissue{11}),
\bfpage{113001}
(\byear{2015})
\end{barticle}
\endbibitem

\bibitem[\protect\citeauthoryear{J{\"a}ckel et~al.}{2010}]{jackel2010all}
\begin{barticle}
\bauthor{\bsnm{J{\"a}ckel}, \binits{O.}},
\bauthor{\bsnm{Polz}, \binits{J.}},
\bauthor{\bsnm{Pfotenhauer}, \binits{S.}},
\bauthor{\bsnm{Schlenvoigt}, \binits{H.}},
\bauthor{\bsnm{Schwoerer}, \binits{H.}},
\bauthor{\bsnm{Kaluza}, \binits{M.}}:
\batitle{All-optical measurement of the hot electron sheath driving laser ion acceleration from thin foils}.
\bjtitle{New Journal of Physics}
\bvolume{12}(\bissue{10}),
\bfpage{103027}
(\byear{2010})
\end{barticle}
\endbibitem

\bibitem[\protect\citeauthoryear{Macchi et~al.}{2013}]{macchi2013advanced}
\begin{barticle}
\bauthor{\bsnm{Macchi}, \binits{A.}},
\bauthor{\bsnm{Sgattoni}, \binits{A.}},
\bauthor{\bsnm{Sinigardi}, \binits{S.}},
\bauthor{\bsnm{Borghesi}, \binits{M.}},
\bauthor{\bsnm{Passoni}, \binits{M.}}:
\batitle{Advanced strategies for ion acceleration using high-power lasers}.
\bjtitle{Plasma Physics and Controlled Fusion}
\bvolume{55}(\bissue{12}),
\bfpage{124020}
(\byear{2013})
\end{barticle}
\endbibitem

\bibitem[\protect\citeauthoryear{L{\'e}cz et~al.}{2013}]{lecz2013target}
\begin{barticle}
\bauthor{\bsnm{L{\'e}cz}, \binits{Z.}},
\bauthor{\bsnm{Boine-Frankenheim}, \binits{O.}},
\bauthor{\bsnm{Kornilov}, \binits{V.}}:
\batitle{Target normal sheath acceleration for arbitrary proton layer thickness}.
\bjtitle{Nuclear Instruments and Methods in Physics Research Section A: Accelerators, Spectrometers, Detectors and Associated Equipment}
\bvolume{727},
\bfpage{51}--\blpage{58}
(\byear{2013})
\end{barticle}
\endbibitem

\bibitem[\protect\citeauthoryear{DuBois et~al.}{2017}]{dubois2017origins}
\begin{botherref}
\oauthor{\bsnm{DuBois}, \binits{T.C.}},
\oauthor{\bsnm{Siminos}, \binits{E.}},
\oauthor{\bsnm{Ferri}, \binits{J.}},
\oauthor{\bsnm{Gremillet}, \binits{L.}},
\oauthor{\bsnm{F{\"u}l{\"o}p}, \binits{T.}}:
Origins of plateau formation in ion energy spectra under target normal sheath acceleration.
Physics of Plasmas
\textbf{24}(12)
(2017)
\end{botherref}
\endbibitem

\bibitem[\protect\citeauthoryear{Fang et~al.}{2016}]{fang2016combined}
\begin{barticle}
\bauthor{\bsnm{Fang}, \binits{Y.}},
\bauthor{\bsnm{Yu}, \binits{T.}},
\bauthor{\bsnm{Ge}, \binits{X.}},
\bauthor{\bsnm{Yang}, \binits{S.}},
\bauthor{\bsnm{Wei}, \binits{W.}},
\bauthor{\bsnm{Yuan}, \binits{T.}},
\bauthor{\bsnm{Liu}, \binits{F.}},
\bauthor{\bsnm{Chen}, \binits{M.}},
\bauthor{\bsnm{Liu}, \binits{J.}},
\bauthor{\bsnm{Li}, \binits{Y.}}, \betal:
\batitle{Combined proton acceleration from foil targets by ultraintense short laser pulses}.
\bjtitle{Plasma Physics and Controlled Fusion}
\bvolume{58}(\bissue{4}),
\bfpage{045025}
(\byear{2016})
\end{barticle}
\endbibitem

\bibitem[\protect\citeauthoryear{Robinson et~al.}{2006}]{robinson2006effect}
\begin{barticle}
\bauthor{\bsnm{Robinson}, \binits{A.L.}},
\bauthor{\bsnm{Bell}, \binits{A.}},
\bauthor{\bsnm{Kingham}, \binits{R.}}:
\batitle{Effect of target composition on proton energy spectra in ultraintense laser-solid interactions}.
\bjtitle{Physical review letters}
\bvolume{96}(\bissue{3}),
\bfpage{035005}
(\byear{2006})
\end{barticle}
\endbibitem

\bibitem[\protect\citeauthoryear{Borghesi et~al.}{2004}]{borghesi2004multi}
\begin{barticle}
\bauthor{\bsnm{Borghesi}, \binits{M.}},
\bauthor{\bsnm{Mackinnon}, \binits{A.}},
\bauthor{\bsnm{Campbell}, \binits{D.H.}},
\bauthor{\bsnm{Hicks}, \binits{D.}},
\bauthor{\bsnm{Kar}, \binits{S.}},
\bauthor{\bsnm{Patel}, \binits{P.K.}},
\bauthor{\bsnm{Price}, \binits{D.}},
\bauthor{\bsnm{Romagnani}, \binits{L.}},
\bauthor{\bsnm{Schiavi}, \binits{A.}},
\bauthor{\bsnm{Willi}, \binits{O.}}:
\batitle{Multi-mev proton source investigations in ultraintense laser-foil interactions}.
\bjtitle{Physical Review Letters}
\bvolume{92}(\bissue{5}),
\bfpage{055003}
(\byear{2004})
\end{barticle}
\endbibitem

\bibitem[\protect\citeauthoryear{Roth and Schollmeier}{2017}]{roth2017ion}
\begin{botherref}
\oauthor{\bsnm{Roth}, \binits{M.}},
\oauthor{\bsnm{Schollmeier}, \binits{M.}}:
Ion acceleration-target normal sheath acceleration.
arXiv preprint arXiv:1705.10569
(2017)
\end{botherref}
\endbibitem

\bibitem[\protect\citeauthoryear{Mora}{2003}]{mora2003plasma}
\begin{barticle}
\bauthor{\bsnm{Mora}, \binits{P.}}:
\batitle{Plasma expansion into a vacuum}.
\bjtitle{Physical Review Letters}
\bvolume{90}(\bissue{18}),
\bfpage{185002}
(\byear{2003})
\end{barticle}
\endbibitem

\bibitem[\protect\citeauthoryear{Fuchs et~al.}{2006}]{fuchs2006laser}
\begin{barticle}
\bauthor{\bsnm{Fuchs}, \binits{J.}},
\bauthor{\bsnm{Antici}, \binits{P.}},
\bauthor{\bsnm{d’Humi{\`e}res}, \binits{E.}},
\bauthor{\bsnm{Lefebvre}, \binits{E.}},
\bauthor{\bsnm{Borghesi}, \binits{M.}},
\bauthor{\bsnm{Brambrink}, \binits{E.}},
\bauthor{\bsnm{Cecchetti}, \binits{C.}},
\bauthor{\bsnm{Kaluza}, \binits{M.}},
\bauthor{\bsnm{Malka}, \binits{V.}},
\bauthor{\bsnm{Manclossi}, \binits{M.}}, \betal:
\batitle{Laser-driven proton scaling laws and new paths towards energy increase}.
\bjtitle{Nature physics}
\bvolume{2}(\bissue{1}),
\bfpage{48}--\blpage{54}
(\byear{2006})
\end{barticle}
\endbibitem

\bibitem[\protect\citeauthoryear{Johnson}{2017}]{johnson2017review}
\begin{barticle}
\bauthor{\bsnm{Johnson}, \binits{R.P.}}:
\batitle{Review of medical radiography and tomography with proton beams}.
\bjtitle{Reports on progress in physics}
\bvolume{81}(\bissue{1}),
\bfpage{016701}
(\byear{2017})
\end{barticle}
\endbibitem

\bibitem[\protect\citeauthoryear{Wang et~al.}{2015}]{wang2015large}
\begin{botherref}
\oauthor{\bsnm{Wang}, \binits{W.}},
\oauthor{\bsnm{Shen}, \binits{B.}},
\oauthor{\bsnm{Zhang}, \binits{H.}},
\oauthor{\bsnm{Lu}, \binits{X.}},
\oauthor{\bsnm{Wang}, \binits{C.}},
\oauthor{\bsnm{Liu}, \binits{Y.}},
\oauthor{\bsnm{Yu}, \binits{L.}},
\oauthor{\bsnm{Chu}, \binits{Y.}},
\oauthor{\bsnm{Li}, \binits{Y.}},
\oauthor{\bsnm{Xu}, \binits{T.}}, et al.:
Large-scale proton radiography with micrometer spatial resolution using femtosecond petawatt laser system.
AIP Advances
\textbf{5}(10)
(2015)
\end{botherref}
\endbibitem

\bibitem[\protect\citeauthoryear{Li et~al.}{2021}]{li2021influence}
\begin{botherref}
\oauthor{\bsnm{Li}, \binits{D.}},
\oauthor{\bsnm{Xu}, \binits{X.}},
\oauthor{\bsnm{Yang}, \binits{T.}},
\oauthor{\bsnm{Wu}, \binits{M.}},
\oauthor{\bsnm{Zhang}, \binits{Y.}},
\oauthor{\bsnm{Cheng}, \binits{H.}},
\oauthor{\bsnm{Hu}, \binits{X.}},
\oauthor{\bsnm{Geng}, \binits{Y.}},
\oauthor{\bsnm{Zhu}, \binits{J.}},
\oauthor{\bsnm{Zhao}, \binits{Y.}}, et al.:
Influence factors of resolution in laser accelerated proton radiography and image deblurring.
AIP Advances
\textbf{11}(8)
(2021)
\end{botherref}
\endbibitem

\bibitem[\protect\citeauthoryear{Pfotenhauer et~al.}{2010}]{pfotenhauer2010cascaded}
\begin{barticle}
\bauthor{\bsnm{Pfotenhauer}, \binits{S.}},
\bauthor{\bsnm{J{\"a}ckel}, \binits{O.}},
\bauthor{\bsnm{Polz}, \binits{J.}},
\bauthor{\bsnm{Steinke}, \binits{S.}},
\bauthor{\bsnm{Schlenvoigt}, \binits{H.}},
\bauthor{\bsnm{Heymann}, \binits{J.}},
\bauthor{\bsnm{Robinson}, \binits{A.}},
\bauthor{\bsnm{Kaluza}, \binits{M.}}:
\batitle{A cascaded laser acceleration scheme for the generation of spectrally controlled proton beams}.
\bjtitle{New Journal of Physics}
\bvolume{12}(\bissue{10}),
\bfpage{103009}
(\byear{2010})
\end{barticle}
\endbibitem

\bibitem[\protect\citeauthoryear{Wang et~al.}{2018}]{wang2018multi}
\begin{botherref}
\oauthor{\bsnm{Wang}, \binits{W.}},
\oauthor{\bsnm{Shen}, \binits{B.}},
\oauthor{\bsnm{Zhang}, \binits{H.}},
\oauthor{\bsnm{Lu}, \binits{X.}},
\oauthor{\bsnm{Li}, \binits{J.}},
\oauthor{\bsnm{Zhai}, \binits{S.}},
\oauthor{\bsnm{Li}, \binits{S.}},
\oauthor{\bsnm{Wang}, \binits{X.}},
\oauthor{\bsnm{Xu}, \binits{R.}},
\oauthor{\bsnm{Wang}, \binits{C.}}, et al.:
Multi-stage proton acceleration controlled by double beam image technique.
Physics of Plasmas
\textbf{25}(6)
(2018)
\end{botherref}
\endbibitem

\bibitem[\protect\citeauthoryear{Aurand et~al.}{2016}]{aurand2016manipulation}
\begin{botherref}
\oauthor{\bsnm{Aurand}, \binits{B.}},
\oauthor{\bsnm{Senje}, \binits{L.}},
\oauthor{\bsnm{Svensson}, \binits{K.}},
\oauthor{\bsnm{Hansson}, \binits{M.}},
\oauthor{\bsnm{Higginson}, \binits{A.}},
\oauthor{\bsnm{Gonoskov}, \binits{A.}},
\oauthor{\bsnm{Marklund}, \binits{M.}},
\oauthor{\bsnm{Persson}, \binits{A.}},
\oauthor{\bsnm{Lundh}, \binits{O.}},
\oauthor{\bsnm{Neely}, \binits{D.}}, et al.:
Manipulation of the spatial distribution of laser-accelerated proton beams by varying the laser intensity distribution.
Physics of Plasmas
\textbf{23}(2)
(2016)
\end{botherref}
\endbibitem

\bibitem[\protect\citeauthoryear{Hou et~al.}{2009}]{hou2009mev}
\begin{botherref}
\oauthor{\bsnm{Hou}, \binits{B.}},
\oauthor{\bsnm{Nees}, \binits{J.}},
\oauthor{\bsnm{Easter}, \binits{J.}},
\oauthor{\bsnm{Davis}, \binits{J.}},
\oauthor{\bsnm{Petrov}, \binits{G.}},
\oauthor{\bsnm{Thomas}, \binits{A.}},
\oauthor{\bsnm{Krushelnick}, \binits{K.}}:
Mev proton beams generated by 3 mj ultrafast laser pulses at 0.5 khz.
Applied Physics Letters
\textbf{95}(10)
(2009)
\end{botherref}
\endbibitem

\bibitem[\protect\citeauthoryear{Noaman-ul Haq et~al.}{2017}]{noaman2017statistical}
\begin{barticle}
\bauthor{\bsnm{Noaman-ul-Haq}, \binits{M.}},
\bauthor{\bsnm{Ahmed}, \binits{H.}},
\bauthor{\bsnm{Sokollik}, \binits{T.}},
\bauthor{\bsnm{Yu}, \binits{L.}},
\bauthor{\bsnm{Liu}, \binits{Z.}},
\bauthor{\bsnm{Yuan}, \binits{X.}},
\bauthor{\bsnm{Yuan}, \binits{F.}},
\bauthor{\bsnm{Mirzaie}, \binits{M.}},
\bauthor{\bsnm{Ge}, \binits{X.}},
\bauthor{\bsnm{Chen}, \binits{L.}}, \betal:
\batitle{Statistical analysis of laser driven protons using a high-repetition-rate tape drive target system}.
\bjtitle{Physical Review Accelerators and Beams}
\bvolume{20}(\bissue{4}),
\bfpage{041301}
(\byear{2017})
\end{barticle}
\endbibitem

\bibitem[\protect\citeauthoryear{Morrison et~al.}{2018}]{morrison2018mev}
\begin{barticle}
\bauthor{\bsnm{Morrison}, \binits{J.T.}},
\bauthor{\bsnm{Feister}, \binits{S.}},
\bauthor{\bsnm{Frische}, \binits{K.D.}},
\bauthor{\bsnm{Austin}, \binits{D.R.}},
\bauthor{\bsnm{Ngirmang}, \binits{G.K.}},
\bauthor{\bsnm{Murphy}, \binits{N.R.}},
\bauthor{\bsnm{Orban}, \binits{C.}},
\bauthor{\bsnm{Chowdhury}, \binits{E.A.}},
\bauthor{\bsnm{Roquemore}, \binits{W.}}:
\batitle{Mev proton acceleration at khz repetition rate from ultra-intense laser liquid interaction}.
\bjtitle{New Journal of Physics}
\bvolume{20}(\bissue{2}),
\bfpage{022001}
(\byear{2018})
\end{barticle}
\endbibitem

\bibitem[\protect\citeauthoryear{Chou et~al.}{2022}]{treffert2022high}
\begin{botherref}
\oauthor{\bsnm{Chou}, \binits{H.-G.}}, et al.:
High-repetition-rate, multi-mev deuteron acceleration from converging heavy water microjets at laser intensities of 1021 w/cm2.
Applied Physics Letters
\textbf{121}(7)
(2022)
\end{botherref}
\endbibitem

\bibitem[\protect\citeauthoryear{Willingale et~al.}{2006}]{willingale2006collimated}
\begin{barticle}
\bauthor{\bsnm{Willingale}, \binits{L.}},
\bauthor{\bsnm{Mangles}, \binits{S.}},
\bauthor{\bsnm{Nilson}, \binits{P.}},
\bauthor{\bsnm{Clarke}, \binits{R.}},
\bauthor{\bsnm{Dangor}, \binits{A.}},
\bauthor{\bsnm{Kaluza}, \binits{M.}},
\bauthor{\bsnm{Karsch}, \binits{S.}},
\bauthor{\bsnm{Lancaster}, \binits{K.}},
\bauthor{\bsnm{Mori}, \binits{W.}},
\bauthor{\bsnm{Najmudin}, \binits{Z.}}, \betal:
\batitle{Collimated multi-mev ion beams from high-intensity laser interactions<? format?> with underdense plasma}.
\bjtitle{Physical review letters}
\bvolume{96}(\bissue{24}),
\bfpage{245002}
(\byear{2006})
\end{barticle}
\endbibitem

\bibitem[\protect\citeauthoryear{Fukuda et~al.}{2009}]{fukuda2009energy}
\begin{barticle}
\bauthor{\bsnm{Fukuda}, \binits{Y.}},
\bauthor{\bsnm{Faenov}, \binits{A.Y.}},
\bauthor{\bsnm{Tampo}, \binits{M.}},
\bauthor{\bsnm{Pikuz}, \binits{T.}},
\bauthor{\bsnm{Nakamura}, \binits{T.}},
\bauthor{\bsnm{Kando}, \binits{M.}},
\bauthor{\bsnm{Hayashi}, \binits{Y.}},
\bauthor{\bsnm{Yogo}, \binits{A.}},
\bauthor{\bsnm{Sakaki}, \binits{H.}},
\bauthor{\bsnm{Kameshima}, \binits{T.}}, \betal:
\batitle{Energy increase in multi-mev ion acceleration in the interaction of a short pulse laser with a cluster-gas target}.
\bjtitle{Physical review letters}
\bvolume{103}(\bissue{16}),
\bfpage{165002}
(\byear{2009})
\end{barticle}
\endbibitem

\bibitem[\protect\citeauthoryear{Cohen et~al.}{2024}]{cohen2024accumulated}
\begin{barticle}
\bauthor{\bsnm{Cohen}, \binits{I.}},
\bauthor{\bsnm{Cohen}, \binits{T.}},
\bauthor{\bsnm{Levinson}, \binits{A.}},
\bauthor{\bsnm{Elkind}, \binits{M.}},
\bauthor{\bsnm{Rakovsky}, \binits{Y.}},
\bauthor{\bsnm{Levanon}, \binits{A.}},
\bauthor{\bsnm{Michaeli}, \binits{D.}},
\bauthor{\bsnm{Cohen}, \binits{E.}},
\bauthor{\bsnm{Beck}, \binits{A.}},
\bauthor{\bsnm{Pomerantz}, \binits{I.}}:
\batitle{Accumulated laser-photoneutron generation}.
\bjtitle{The European Physical Journal Plus}
\bvolume{139}(\bissue{7}),
\bfpage{1}--\blpage{7}
(\byear{2024})
\end{barticle}
\endbibitem

\bibitem[\protect\citeauthoryear{Kawahito and Kishimoto}{2020}]{kawahito2020ionization}
\begin{botherref}
\oauthor{\bsnm{Kawahito}, \binits{D.}},
\oauthor{\bsnm{Kishimoto}, \binits{Y.}}:
Ionization and acceleration of multiply charged gold ions in solid film irradiated by high intensity laser.
Physics of Plasmas
\textbf{27}(3)
(2020)
\end{botherref}
\endbibitem

\bibitem[\protect\citeauthoryear{Cantono et~al.}{2021}]{cantono2021laser}
\begin{barticle}
\bauthor{\bsnm{Cantono}, \binits{G.}},
\bauthor{\bsnm{Permogorov}, \binits{A.}},
\bauthor{\bsnm{Ferri}, \binits{J.}},
\bauthor{\bsnm{Smetanina}, \binits{E.}},
\bauthor{\bsnm{Dmitriev}, \binits{A.}},
\bauthor{\bsnm{Persson}, \binits{A.}},
\bauthor{\bsnm{F{\"u}l{\"o}p}, \binits{T.}},
\bauthor{\bsnm{Wahlstr{\"o}m}, \binits{C.-G.}}:
\batitle{Laser-driven proton acceleration from ultrathin foils with nanoholes}.
\bjtitle{Scientific Reports}
\bvolume{11}(\bissue{1}),
\bfpage{5006}
(\byear{2021})
\end{barticle}
\endbibitem

\bibitem[\protect\citeauthoryear{Kluge et~al.}{2010}]{Kluge2010mass-limited-foils}
\begin{barticle}
\bauthor{\bsnm{Kluge}, \binits{T.}},
\bauthor{\bsnm{Enghardt}, \binits{W.}},
\bauthor{\bsnm{Kraft}, \binits{S.D.}},
\bauthor{\bsnm{Schramm}, \binits{U.}},
\bauthor{\bsnm{Zeil}, \binits{K.}},
\bauthor{\bsnm{Cowan}, \binits{T.E.}},
\bauthor{\bsnm{Bussmann}, \binits{M.}}:
\batitle{{Enhanced laser ion acceleration from mass-limited foils}}.
\bjtitle{Physics of Plasmas}
\bvolume{17}(\bissue{12}),
\bfpage{123103}
(\byear{2010})
\doiurl{10.1063/1.3519512}
\end{barticle}
\endbibitem

\bibitem[\protect\citeauthoryear{Porat et~al.}{2021}]{PhysRevResearch.3.L032059}
\begin{barticle}
\bauthor{\bsnm{Porat}, \binits{E.}},
\bauthor{\bsnm{Yehuda}, \binits{H.}},
\bauthor{\bsnm{Cohen}, \binits{I.}},
\bauthor{\bsnm{Levanon}, \binits{A.}},
\bauthor{\bsnm{Pomerantz}, \binits{I.}}:
\batitle{Diffraction-limited coherent wake emission}.
\bjtitle{Phys. Rev. Research}
\bvolume{3},
\bfpage{032059}
(\byear{2021})
\doiurl{10.1103/PhysRevResearch.3.L032059}
\end{barticle}
\endbibitem

\bibitem[\protect\citeauthoryear{Pappalardo et~al.}{2010}]{Pappalardo2010}
\begin{botherref}
\oauthor{\bsnm{Pappalardo}, \binits{A.}},
\oauthor{\bsnm{Cosentino}, \binits{L.}},
\oauthor{\bsnm{Finocchiaro}, \binits{P.}}:
{An imaging technique for detection and absolute calibration of scintillation light}.
Review of Scientific Instruments
\textbf{81}(3)
(2010)
\doiurl{10.1063/1.3360931}
\end{botherref}
\endbibitem

\bibitem[\protect\citeauthoryear{Cohen et~al.}{2024}]{cohen2024undepleted}
\begin{barticle}
\bauthor{\bsnm{Cohen}, \binits{I.}},
\bauthor{\bsnm{Meir}, \binits{T.}},
\bauthor{\bsnm{Tangtartharakul}, \binits{K.}},
\bauthor{\bsnm{Perelmutter}, \binits{L.}},
\bauthor{\bsnm{Elkind}, \binits{M.}},
\bauthor{\bsnm{Gershuni}, \binits{Y.}},
\bauthor{\bsnm{Levanon}, \binits{A.}},
\bauthor{\bsnm{Arefiev}, \binits{A.V.}},
\bauthor{\bsnm{Pomerantz}, \binits{I.}}:
\batitle{Undepleted direct laser acceleration}.
\bjtitle{Science Advances}
\bvolume{10}(\bissue{2}),
\bfpage{1947}
(\byear{2024})
\end{barticle}
\endbibitem

\bibitem[\protect\citeauthoryear{Ziegler et~al.}{2008}]{srim}
\begin{botherref}
\oauthor{\bsnm{Ziegler}, \binits{F.}},
\oauthor{\bsnm{Biersack}, \binits{J.}},
\oauthor{\bsnm{Ziegler}, \binits{M.}}:
Srim--the stopping and range of ions in solids srim co.
Chester, MD
(2008)
\end{botherref}
\endbibitem

\bibitem[\protect\citeauthoryear{Man{\v{c}}i{\'c} et~al.}{2008}]{manvcic2008absolute}
\begin{botherref}
\oauthor{\bsnm{Man{\v{c}}i{\'c}}, \binits{A.}},
\oauthor{\bsnm{Fuchs}, \binits{J.}},
\oauthor{\bsnm{Antici}, \binits{P.}},
\oauthor{\bsnm{Gaillard}, \binits{S.}},
\oauthor{\bsnm{Audebert}, \binits{P.}}:
Absolute calibration of photostimulable image plate detectors used as (0.5--20mev) high-energy proton detectors.
Review of Scientific Instruments
\textbf{79}(7)
(2008)
\end{botherref}
\endbibitem

\end{thebibliography}

\section*{Acknowledgments}
This research was supported by Israel Science Foundation Grant No. 2314/21.
We acknowledge the aid in the fabrication of the targets from the Tel Aviv University Center for Nanoscience and Nanotechnology. 
Simulations were performed using EPOCH, which was developed
as part of the UK Engineering and Physical Sciences Research Council (EPSRC)-funded Project No. EP/G054940/1.
I.P. acknowledges the support of the Zuckerman STEM Leadership Program.
We acknowledge the EuroHPC Joint Undertaking for awarding this project access to the EuroHPC supercomputer LUMI, hosted by CSC (Finland) and the LUMI consortium.

\section*{Author contributions}
I.P. conceptualized, obtained funding and supervised this project. 
M.E. performed and analyzed the experiment, with assistance from N.P., I.C., N.A. A.Levi. and A.Leva..
M.E. and I.P. performed and analyzed the PIC simulations,
and wrote the manuscript with feedback from all other authors. 
\section*{Competing interests}
The authors declare no competing interests.

\end{document}